\documentclass[11pt]{article}

\usepackage{amsmath, amssymb}
 \usepackage{epsfig}
\def \td {\tilde}
\def \G {\Gamma }

\newcommand \be{\begin{eqnarray}}
\newcommand \ee{\end{eqnarray}}

\newcommand{\ba}{\begin{eqnarray}}
\newcommand{\ea}{\end{eqnarray}}
\newcommand{\oh}[1]{{\cal O}( #1 )}
\newcommand{\refeq}[1]{Eq.~(\ref{eq:#1})}

\def\sl{\sqrt{\lambda}}
\def\l {\lambda}
\def \g{\gamma}
\def\o{\omega}
\def\S{{\cal S}}
\def\r{\rho}
\def\a{\alpha}
\def\k{\kappa}
\def\E{{\cal E}}
\def\n{\nu}

\def\s{\sigma}

 \def \tS{\bar {\cal S}}

 \def\th{\theta}
 \def\C{{\cal C}}
 \def\hC{{\hat \C}}
 \def\tg{\tilde{\g}}
 \def\tP{\tilde{\P}}
 \def\hS{{\hat \S}}
 \def\J{{\cal J}}
\def \adss  {$AdS_5 \times S^5$}
\def \sql {\sqrt{\lambda}}
\def\O{{\cal O}}

\begin{document}


\renewcommand{\thefootnote}{\arabic{footnote}}


\def \foot {\footnote}
\def \bi{\bibitem}

\def \tr {{\rm tr}}
\def \ha {{1 \over 2}}

\def \ci{\cite}
\def \N {{\mathcal N}}
\def \ww {\Omega}
\def \const {{\rm const}}
\def \t {\tau}
\def\S{{\mathcal S} }
\def \nn {\nu}
\def \XX {{\rm X}}

\def \k {\kappa}
\def\foot{\footnote}
\def \four{{\textstyle {1\ov 4}}}
 \def \third { \textstyle {1\ov 3
}}
\def\det{\hbox{det}}
\def \ci {\cite}
\def \ov {\over}

\def \bp {\begin{pmatrix}}  \def \epm {\end{pmatrix}}
\def \ha {{\textstyle{1 \ov 2}}}

\def \bi{\bibitem}
\def \la {\label}
\def \Tr  {{\rm Tr}}

\def \T {{\cal T}}
\def \l {\lambda}
\def\foot{\footnote}
\def \tl  {{\tilde \l}}
\def \sql {{\sqrt \l}}
\def \adss {$AdS_5 \times S^5$\ }
\newcommand{\rf}[1]{(\ref{#1})}

\def \qr {{\hat \rho}}

\overfullrule=0pt
\parskip=2pt
\parindent=12pt
\headheight=0in \headsep=0in \topmargin=0in \oddsidemargin=0in

\vspace{ -3cm} \thispagestyle{empty} \vspace{-1cm}
\begin{flushright} AEI-2008-076
\end{flushright}
 \vspace{-1cm}
\begin{flushright} HU-EP-08/31
\end{flushright}
 \vspace{-1cm}
\begin{flushright} Imperial-TP-AT-2008-4
\end{flushright}
\begin{center}
{\Large\bf
 Structure of large spin expansion of anomalous dimensions \\
 at strong coupling
 }

 \vspace{.8cm} {
  M. Beccaria$^{a,}$\footnote{matteo.beccaria@le.infn.it}, V.  Forini$^{b,}$\footnote{
  Recently moved to   Albert-Einstein-Institut, MPG, Potsdam.
   forini@aei.mpg.de   },
 A. Tirziu$^{c,}$\footnote{atirziu@purdue.edu} and A.A.
 Tseytlin$^{d,}$\footnote{Also at
 Lebedev  Institute, Moscow.\ \
  tseytlin@imperial.ac.uk
 }}\\
 \vskip 0.1cm

{\em
$^{a}$
Physics Department, Salento University and INFN, 73100 Lecce, Italy \\
\vskip 0.04cm
$^{b}$ Humboldt-Universit\"at zu Berlin,
Institut f\"ur Physik,
 D-12489 Berlin, Germany
   \\
 \vskip 0.08cm
$^{c}$
Department of Physics, Purdue  University,
W. Lafayette, IN 47907-2036, USA\\
\vskip 0.04cm
$^{d}$  The Blackett Laboratory, Imperial College,
London SW7 2AZ, U.K. }

\end{center}

 \vskip 0.8cm

 \begin{abstract}
The anomalous dimensions  of planar $\N=4$   SYM  theory operators like  $\tr( \Phi D_+ ^S \Phi)$
expanded  in  large spin $S$  have the asymptotics $\gamma=
f \ln S + f_c + {1 \ov S} (f_{11} \ln S + f_{10}) + ... $,
where    $f$  (the universal scaling function or cusp anomaly), $f_c$
and $f_{mn}$ are given by power series in the `t Hooft coupling $\l$.
The subleading coefficients appear  to be related by the
 so called functional relation
 and parity (reciprocity) property of the function expressing $\g$  in terms
 of the conformal spin of the collinear group.
 Here we study the structure of such large spin expansion at strong coupling  via AdS/CFT,
 i.e. by using the dual description in terms of folded spinning string in $AdS_5$.
The large spin expansion of the classical  string energy
happens to have  exactly   the same structure as that of  $\g$  in the
perturbative gauge theory.
Moreover, the functional  relation and the reciprocity constraints on
 the coefficients   are also satisfied. We compute
  the leading string 1-loop   corrections to the  coefficients
 $f_c,f_{11},f_{10}$  and verify  the functional/reciprocity relations  at
 subleading $1 \ov \sqrt { \lambda}$ order.
This provides a strong indication that these relations hold not only in weak coupling
 (gauge-theory)  but also in  strong  coupling (string-theory) perturbative expansions.
\end{abstract}
\newpage

\def \no {\nonumber}
\def \bea {\be}
\def \eea {\ee}

\def \PP  {{\rm f}}
\def \P {{\PP}}

\def \ed {\end{document}}
\def \la {\label}

\def \cs {{\sigma}}

\def \Pa {{\rm F}}
\newcommand{\ellK}{\mathbb{K}}
\newcommand{\ellPi}{{\mathnormal{\Pi}}}
\newcommand{\ellE}{\mathbb{E}}

\def \tt {{\rm t}}

\def \n {{\cal J}}
\def \tS {\bar \S}
\def \ka {\kappa_0}

\def \EE {{\ellE}}
\def \KK {{\ellK}}
\def \DD {{\rm D}}
\def \fw {{ \tilde f_c}}
\def \OO  {{\cal O}}

\def \Si {\Sigma}

\def \bS {\bar S}
\def \no  {\nonumber}
\def \F {{_F}}
\def \ff {{\rm r}}
\def \ck {{\rm c }}

\renewcommand{\theequation}{1.\arabic{equation}}
 \setcounter{equation}{0}

\setcounter{equation}{0}
\setcounter{footnote}{0}
\setcounter{section}{0}

\section{Introduction and summary}

Recent advances  in the study of duality between
planar $\N=4$  SYM theory and   free \adss superstring theory which
utilise their integrability property
led to important insights into the structure of dependence of anomalous
dimensions of gauge-invariant operators on the quantum numbers like spin
and  on the t'Hooft  coupling. While  there was a remarkable recent progress 
in understanding the asymptotic large spin limit
 in which
the compactness of the spatial direction of the world sheet may be ignored
 \ci{bes,frs},\foot{
Here we refer to the integral equations that describe the minimal 
anomalous dimension  in the band \ci{bgk,bkp}.  These
equations were obtained from the all-loop Bethe ansatz by taking a
special scaling limit \ci{korch,bgk}
 which describes a condensation of magnons and holes
at the origin.} 
 it is  important to study corrections
to this limit.

Here we shall consider  the famous example \ci{gkp}
of  folded spinning string in $AdS_5$  dual to
a minimal twist gauge theory operator like $\tr ( \Phi \DD_+^S \Phi)$.
Starting with the classical string  energy for the solution of \ci{dev,gkp}
and expanding it in large semiclassical spin parameter
$\S$ one  finds   (see \ci{pt} and below)
\be
 && E = \sql\ \E (\S)  \ , \ \ \ \   \ \ \ \
\S = { S \ov \sql} \ , \\
&& \E(\S)_{_{\S \gg 1}} =\S+ a_0 \ln \S + a_c +
{1 \ov \S} (a_{11}  \ln \S + a_{10}) \no \\
&& \ \ \ \ \ \ \ \ \ \ \ \ \ \ \ \ \ \ \  + \
{1 \ov \S^2} (a_{22}  \ln^2 \S + a_{21} \ln \S + a_{20} ) + \OO( {\ln^3 \S \ov \S^3})
\ ,  \la{exx}
 \ee
with
$a_0= { 1 \ov \pi }, \ \ a_c= { 1 \ov \pi } (\ln 8 \pi-1),$ etc.\foot{Note that
 the  small $\S$ behaviour
of the energy is quite different \ci{tt}:  $\E = \sqrt{ 2 \S}[h_0 + h_1 \S + ... ]$.}
That means that in the semiclassical
string theory limit in which one first takes
 the string tension  $\sql \ov 2 \pi $  to be  large for fixed
  $\S$  and then expands in
 large
$\S$,
  the corresponding string energy  can be written as
  ($\sql \gg 1, \ \ {S \ov \sql}  \gg 1 $)
  \be
&& E
  =S+ f \ln S +  f_c +
{1 \ov S} [f_{11}  \ln S  + f_{10}] \no \\
&& \ \ \ \  \ \ \  \ \ \ \  \ \ \   + \
{1 \ov S^2} [f_{22} \ln^2 S  + f_{21} \ln S + f_{20}] +
\OO( {\ln^3 S \ov S^3}) \ ,  \la{xax}
\ee
where $f= \sql\ a_0 + .., \ f_c = \sql\ a_c + ..., $ etc.
The subleading coefficients simplify  if we  absorb the constant $f_c$ into
 the $\ln S$ term, i.e.  if we re-write \rf{xax} as
 \be
&& E
=
S+ f \ln (S/\fw ) +
{1 \ov S} [f_{11}  \ln (S/\fw )  + f'_{10}] \no \\
&& \ \ \ \  \ \ \  \ \ \ \  \ \ \ \ \  + \
{1 \ov S^2} [f_{22} \ln^2 (S/\fw )  + f'_{21} \ln (S/\fw ) + f'_{20}] +
\OO( {\ln^3 S \ov S^3}) \ , \la{gh} \ee
where to leading order in $1 \ov \sql  $ expansion
\be
f=\frac{\sqrt{\l}}{\pi}
\ , \ \ \fw = { e \sql \ov 8 \pi }  , \ \
f_{11}= \frac{\l}{2\pi^2}, \ \ ~f'_{10}=0  , \ \
f_{22}=-\frac{\l^{3/2}}{8\pi^3} , \ \ \
f'_{21}=\frac{5\,\l^{3/2}}{16\pi^3}  , \ \ \
f'_{20}=\frac{\l^{3/2}}{8\pi^3}\no
\ee
Following the analysis of quantum corrections to the folded  string solution in
\ci{ft1}, one may conclude  that this structure of the large $\S$ expansion is preserved
  by the
$\a' \sim { 1 \ov \sql}$ corrections,
with the coefficients $f, f_c, f_{11}, ...$ being promoted to power series in
$1 \ov \sql$, i.e. $f_{mk} \sim \sum_n {b_{mk,n} \ov (\sql)^n}$.

Indeed, as we shall find below,  the 1-loop corrections  for  leading
 coefficients in  \rf{gh}  
 are\foot{The leading correction to $f$  was
  found  in \ci{ft1}.}
  \be
&&f= \frac{\sqrt{\l}}{\pi}\big[  1 - { 3 \ln 2 \ov \sql} + {\cal O} ({ {1 \ov \l }})   \big]\ ,\la{faj}\\
&&
\fw = \frac{e \sqrt{\lambda}}{8 \pi}\big[   1+ \frac{1}{\sqrt{\lambda}}( 3\ln 2   - \ck) 
+{\cal O}  ( {1 \ov \l})\big] \ ,\la{kk} \\
&&
f_{11}= \frac{\lambda}{2 \pi^2} \big[  1- \frac{6 \ln 2}{\sqrt{\lambda}}+ {\cal O} ( {1 \ov \l})\big]\ ,\la{lq} \\
&&
f'_{10}=\frac{\lambda}{2 \pi^2} \big[ 0   -\frac{0}{\sql} +  {\cal O} ( {1 \ov \l})   \big] \ . \la{qi}
   \ee
Here $\ck$  is  a constant that we were not able to determine with the method for evaluation of 
1-loop string correction we used below. 
Equivalently, in \rf{xax} we get the same $f$, $f_{11}$ and
\be
f_c  &=& f \ln { 1\ov \fw}= { \sql  \ov \pi } \Big[\ln {8 \pi\ov \sql} -1
+  { 1 \ov \sql} \big( - 3 \ln 2\ \ln {8 \pi\ov \sql} + \ck\big) + 
  {\cal O} ( {1 \ov \l}) \Big]\ , \la{l} \\
f_{10} &=&  f'_{10}  +  f_{11}\ln { 1\ov \fw} = f'_{10}  +  { f_c \ov f} f_{11}  \la{relp}\\
  &=& \frac{\lambda}{2 \pi^2}\Big[\ln {8 \pi\ov \sql} -1 + { 1 \ov \sql}\big(
   - {6 \ln 2} \big[\ln{8 \pi\ov \sql} - {1 \ov 2}] + \ck \big) 
+ {\cal O} ( {1 \ov \l})\Big] \ . \la{lll}   \ee
The (at first surprising)  vanishing  of the first two terms in  $f'_{10}$ in  \rf{qi} is a consequence of 
the relation £$  f_{10}-  { f_c \ov f} f_{11} =0$ (here verified at 
first two leading orders, i.e.  to  order £ ${\cal O} (\l^0)$ in $  f_{10}$).

In fact, since we also see that  in \rf{faj},\rf{lq} \   $f_{11}= \ha f^2$, 
this relation is equivalent to $£  f_{10}-  \ha  { f_c } f   =0$.
As we shall see below, these relations are  consequences of the ``functional relation''
and  reciprocity   at strong coupling. 
Note that these conditions thus determine 2 our of 4 coefficients 
in the part of $E$ up to order $O({\ln^2 S \ov S^2})$.
We thus led to  expect that $f'_{10}=0$ should  be true in to {\it all orders}   in the strong coupling expansion, 
suggesting the advantage of the form of $E$ in \rf{gh} over \rf{xax}  and the importance 
of the function $\fw$.

Reversing the usual logic, we  may  then conjecture that  structurally
 same large spin  expansion should appear also at
 weak coupling, i.e. in the perturbative expressions for the
corresponding  gauge theory anomalous dimensions.
This is not,  a priori,  guaranteed  since the limit
taken on the gauge theory  side
is   different from the above
 string-theory limit: there one first expands the anomalous dimension
in small  $\l$
at fixed $S$  and then takes $S$  large
in each of the  $\l^n$ coefficients.
Yet,  remarkably,  expanding
in large $S$ the known 
2-, 3- and 4-loop perturbative anomalous  dimensions  of twist 2 and twist 3  operators
  in SYM theory one does find  \ci{kot,moch,pt,dok1,bk,dok2,Kotikov:2007cy,bf} the expression
of the form \rf{xax}   with the  coefficients given by power series in $\l$, i.e.
$f_{mk} \sim  \sum_n a_{mk,n} \l^n$.
\foot{The 4-loop prediction for twist 2  and 3  anomalous dimension 
 at finite $S$  \ci{Kotikov:2007cy} so far
 was not  based on direct gauge theory computation.}

Assuming that the expansion \rf{xax}  or\foot{Here, as in \rf{exx},\rf{xax},
 we suppress the
dependence on finite twist $J$.}
  \be E-S=  \sum_{m=0}^\infty  { e_m (\l, \ln S) \ov S^m}\ , \ \ \ \ \ \ \ \ \ \
  e_m (\l, \ln S)\ = \sum_{k} f_{mk}(\l) \ln^k S    \la{qw} \ee
  applies for any $\l$,
   and given the important role   of
  the  universal scaling function or cusp anomalous dimension $f(\l)$ \ci{kor88}, one may
  raise the question  about the interpretation
  of  other ``interpolating''  functions $f_{mk}(\l)$
   in \rf{xax}.

  From the gauge theory  point of view, the  function $f(\l)$  appears
   in  the asymptotics of anomalous  dimensions  of gauge invariant operators as
   well as
    (for twist 2)
    in the IR asymptotics of gluon scattering amplitudes  related to UV  cusp
   anomaly  of  light-like Wilson loops (for a  review and references see, e.g.,  \ci{ar}).
   On string  side that corresponds,  respectively,  to the  closed string \ci{gkp} and
   the open string \ci{kruz,am} sectors. They are connected
   in the strict large $S$ limit since then the ends of the folded spinning string reach
   the boundary of $AdS_5$ and thus the associated  world surface has
   a  Wilson line interpretation \ci{krtt}.
    This open string sector interpretation  should  not  be expected
    to  apply  to  other subleading coefficients $f_{mk}(\l)$
    since for finite $S$   the end points  of the  folded string no longer
    touch the boundary
    (cf.,  however, \ci{dix,alm}  where the gauge theory  interpretation of the constant $f_c$
    is discussed).

   In fact,   many of the $f_{mk}$ coefficients in  \rf{xax},\rf{qw}   are
   not actually  independent,  as was first observed at few leading orders  in  weak coupling
   expansion and then given a general interpretation  in  \ci{bk,dok2}.
   According to \ci{bk}, these coefficients   are constrained by\ \ (i)
   the so called  ``functional relation''
   suggested  by
   the conformal invariance
    (which  relates the leading $f_{mm}$  functions to
   powers of the scaling  function  $f$ and thus implies their universality)
    and also by\ \ (ii) the ``parity preserving relation'' or
   ``reciprocity''   \ci{moch,bk,dok2,bdm,bf} (which  relates some subleading non-universal
    coefficients, e.g.,  $f_{10}$ to $f_c$, \  $f_{32}$ to $f_{21}$, etc.).

\

   Our aim  here will be to investigate  the presence  of such  relations
   at strong coupling, i.e. in the semiclassical string theory expansion for the spinning
   string states, extending earlier  observations made in \ci{bk}.\foot{Let us note also that
   the  fine structure
    of the constant term  $f_c$ in \rf{xax}
    for ``non-minimal'' operators and  the dual string     states  was  studied
    in \ci{bgk,bkp} and in \ci{dor}.}

\

Before proceeding let us add an important   clarification. 
While for low twists $J=2,3$ in weak-coupling  expansion 
 the powers of $\ln S$ in $e_m$ in the anomalous dimension  \rf{qw} appear to be 
 positive,\foot{Here we  have in mind the {\it minimal} anomalous dimension in the band 
 \ci{bgk}; the ${J^3 \ov \ln^2 S}$ terms appear \ci{ggg} in non-minimal  dimensions  even for 
 low twists. We thank G. Korchemsky for this clarification.}
 for  $J > 3$  one finds also terms with negative powers  like 
  $J^k\ov \ln^m S$.
  Namely, for $J > 3$ the leading term in   $E-S-J$  appears to be
 $e_0 = k_1(J) +   {k_2 (J) \ov \ln^2 S} + ...$, 
 where for  $k_1 = k_{11} J + k_{10},\  \ k_2= k_{23}  J^3  +  k_{22}  J^2  +  
 k_{21}  J  +   k_{20}$   and   $k_2$ vanishes for $J=2,3$.\foot{We thank A. Rej 
 for a discussion of the structure of these terms at 1-loop order in $sl(2)$   sector 
 (see also \ci{bgk}).} 
 The presence of such terms  was first  observed in \ci{bgk} (in the 1-loop approximation
  in the $sl(2)$ sector)
  in the limit 
 of large $J$ and $S$ with $j\equiv {J \ov \ln S} $   
  fixed and small,  i.e. $e_0 = (  k_{11} j + k_{23} j^3 + ...) \ln S $, 
  and they are likely to be present  also for finite  $J$ 
   with $S\gg 1 $.\foot{There is a numerical evidence   for the presence of 
   ${ k_2(J) \ov \ln^2 S}$ term for  $J\geq 4$ from the  analysis 
    of the corresponding 1-loop Baxter equation 
    (A. Rej, private communication). For a general method to derive higher order terms in 
    $1/S$  expansion at fixed $J$   see  \ci{bc}.
    Let us mention also that coefficients in large $S$ expansion
 beyond  cusp anomaly one may be also  controlled by integral equations like the 
 BES \ci{bes}  one (see in this connection \ci{fio}).}
  At strong coupling, i.e. in the string-theory semiclassical expansion, one cannot distinguish 
  between  finite  values of $J$   and $J=0$;  for  large $J$ 
  one finds (see \ci{bgk} and section 2.2  below) 
  that in  a similar limit of large $S$ 
 with   $ \ell  \equiv {J \ov \sql \ln S} $    fixed 
 $e_0 = ( n_1 \ell^2 +  n_2 \ell^4 + ...) \ln S $. Thus the dependence on $J$ and $S$ 
 is different at strong and weak  coupling and to relate the two expansions  one needs a
 non-trivial resummation rather than simple interpolation of coefficients in $\l$ \ci{rt2,frs}.
 
\bigskip

Let us now   review in more detail
what is known at weak coupling  (see \ci{bk} and references there).
The results of explicit higher-loop planar gauge-theory
 computations
of anomalous dimensions $\g(S,J,\l)$  of operators like $\tr ( \DD_+^S \Phi^J)$
($S$ is the Lorentz spin and $J$ is the twist)
were interpreted in \ci{bk}  in the following way (see also \ci{dok2}).
 Observing that such Wilson-type  operators can be classified according to
  representations of the collinear $SL(2, R)$ subgroup of the
   $SO(2,4)$ conformal  group  \ci{bra}  which are labeled by the
    \emph{conformal spin}
     $s=\textstyle{\frac{1}{2}}(S+\Delta)$
      one may argue that the anomalous dimension $\g=\Delta-S-J$  should be a function
     of  $S$ only through its dependence on the conformal spin $s$.
     Since the scaling dimension  $\Delta$
     is\foot{Here we assume that $\Phi$ in the operator $\tr ( \DD_+^S \Phi^J)$
      is a scalar field (as is the case in the $sl(2)$ sector of SYM theory).
     The relation between the notation used in~\cite{bk} and ours  is:  $N\to S$,
       $L\to J$,  $J\to C$ and $j\to  s$.}
     \be \la{del}\Delta = S + J + \g(S,J)\ , \ee  that then leads to the following
     {\it ``functional relation''}
      for $\g$
     \be \g(S,J) =\PP(s; J)= \PP \big( S + \ha J + \ha \g(S,J); J\big) \ .
     \la{ko} \ee
Without further information, this relation is nothing more than a change of variable, 
 since,
at least in perturbation theory, it is always possible to compute the
 function $\PP$ in terms of  the anomalous
dimension $\g(S, J)$. Nevertheless, the above reasoning suggests that $\PP$ could be more fundamental 
than $\g$.\foot{Apart from conformal invariance, for twist 2  case,  
this conclusion is also expected on the basis of the 
QCD origin of the functional relation \rf{ko} \cite{dok1}.
In that context, 
the function $\PP$ for twist 2 operators 
 is closely related to a special reformulation of the parton 
distribution functions evolution equation which aims at treating symmetrically the 
space-like channel of deep inelastic scattering and the time-like crossed channel describing
$e^+e^-$ annihilation \cite{Alt}. In particular, the functional relation \rf{ko} turns out to be
predictive within the large $S$ expansion \rf{qw} because the function
$\PP$ happens to be {\it simpler} than $\gamma$ and, in particular, it does not contain $\ln^m S/S^m$ terms
  \cite{dok2,bk,bdm}.}
      This is what we shall assume below when referring to the  functional relation.

     Suppressing the dependence on $J$ in $\g$ and $\PP$ we may write this functional
     relation simply as\foot{In the approach of \ci{dok1,dok2}, which  recently
     received a nice confirmation in~\cite{Laenen:2008ux},  this relation follows
     from a suitable modification of the evolution equations governing the renormalization
     of the twist operators.}
     \be \g(S) = \PP \big( S + \ha \g(S)\big)  \la{fa} \ .  \ee
At weak coupling $\g(S) = \sum^\infty_{n=1} \g_n(S) \l^n $;  expanding
the coefficients $\g_n$ in  large $S$ (for fixed $J$)   one finds
that for  all explicitly known  perturbative gauge-theory results one  gets the same
expansion as in \rf{xax}
\bea\nonumber
\gamma(S)_{_{S \gg1 }}  &=& f\,\ln\,S
+ f_c + \frac{f_{11}\,\ln\,S+f_{10}}{S} + \frac{f_{22}\,\ln^2\,S+f_{21}\,\ln\,S+f_{20}}{S^2}+\\
&& \ \ \ \ \ \ \ \ \ +\  \frac{f_{33}\,\ln^3\,S + f_{32}\,\ln^2\,S+f_{31}\,\ln\,S+
f_{30}}{S^3}+{\cal O}\big({\ln^4\,S\ov S^4}\big),  \la{xa}
\eea
where the  coefficients $f,f_c,f_{11},...$ are power series in $\l$.
Remarkably, the structure of this expansion turns out to be perfectly consistent
with the functional relation \rf{fa}:
  the function $\PP$ starts with
 a logarithmic term  (and is ``simpler'' than $f$, i.e. has no $\ln^n S \over S^n$ terms in it)
  so that  the coefficients of the leading $
  \ln^m S \ov S^m$ terms are all determined by the scaling function  $f$
  ~\cite{bk,bdm,bkp}
\bea\label{lea}
 \g(S)=f \ln\big(S+\ha f \ln S+ ...\big)+...=
f \ln S+\frac{f^2}{2}\,\frac{\ln S}{S}-\frac{f^3}{8}\,\frac{\ln^2 S}{S^2}+\frac{f^4}{24}\,\frac{
 \ln^3 S}{S^3} + ... \ .
\eea
The universality (i.e. twist and flavor independence)
of the scaling function or cusp anomalous dimension $f$ thus  implies the
universality of all of the coefficients $f_{mm}$  in \rf{xa}
as  they are simply proportional to   $f^{m+1}$,
\be \la{fun}
f_{11} = {1 \ov 2}  f^2, \ \ \  \ \ f_{22} = - {1 \ov 8} f^3 , \ \ \  \ \
f_{33}
=  {1 \ov 24} f^4, \ \ ... \ .  \ee
These   should be understood as relations between  the functions
$f_{mm}(\l)$ and $f(\l)$ defined  as  power series in $\l$.

Let us note that anomalous
dimensions of operators with twist higher than two occupy a band~\cite{bgk},
 the lower bound of which is the  \emph{minimal} dimension for given $S$ and $J$.
 The relation \rf{lea} is expected to apply for the minimal dimension in the band.
Interestingly, as was found  at weak coupling,  a similar   relation also holds for the excited 
trajectories \ci{bkp}. This is also what we shall see at strong coupling 
on the example of the spiky string in section 2.3 (see \rf{spk}).

In Appendix F below
  we will summarize  the known
perturbative expansions  for the minimal  anomalous dimensions
of twist $2$   and twist $3$ operators of various flavors,
as obtained from the asymptotic Bethe ansatz of~\cite{Beisert:2005fw}.
These results are indeed consistent with \rf{lea},\rf{fun}  and thus
 with the universality of the $f_{mm}$ coefficients  in \rf{xa}.

\bigskip

As for the subleading  (${\ln^k S \ov S^m}, \   k < m $) terms in \rf{xa},
their coefficients are,   at least
partially,  controlled  by special properties of the function $\PP$ in
the functional relation \rf{fa}. Indeed, it was observed on many examples
 that the function  $\PP$ should  satisfy a
``parity preserving relation''  or  \emph{reciprocity}
 property.~\footnote{The name reciprocity
has its origin in the 
 the formulation  of this property, for twist 2  case,   in terms of   the  Mellin  transform:
   \ $F(x)=-x\,F(1/x)$, where $\PP(S)=\int_0^1dx\,x^{S-1}\,F(x)$.
   For twist larger than 2 the ``parity preserving relation'' was suggested in \ci{bk}
   as a more appropriate name.  Here for simplicity   we shall not make  this distinction 
   and will often  refer to reciprocity when implying the ``parity invariance''
   property in \rf{rec},\rf{casimir} below.
   }
  This property   implies   that
 the  large $S$ expansion of $\PP(S)$   should   run
  in the inverse {\it even} powers of the
  quadratic Casimir   of  the collinear  $SL(2, R)$ group,  namely,  \ci{bk}
   \be\label{rec}
{\PP}(S)=\sum^\infty_{n=0}\frac{a_n(\ln C)}{C^{2n}} \ ,
\ee
where $C$ is the ``bare''
quadratic Casimir defined in terms of the
``canonical''  value  of the conformal spin $s_0$
as $C^2 \equiv  s_0 (s_0 -1), \ \ s_0=\ha (S + \Delta_0)=
S   +   \ha  J  \ $, i.e.\foot{Here we again consider the operator  built  of scalar fields.
For twist 2, i.e. $J=2$,  one then has $C^2= S(S+1)$, so that $C=S + \ha + \OO({1 \ov S})$.
  For generic flavour  one is to replace, in (\ref{casimir}), $S + \ha J \to S +  \ell J$, where
$\ell=  {\ha}, 1 , {3 \ov 2}$ for a scalar, spinor or  vector cases~\cite{bra}.}
\be
\label{casimir}
C^2 =    (S   +   \ha  J)( S   +   \ha  J  -1)  \ .
\ee
The reciprocity property \rf{rec} of the function  $\PP$ in the relation \rf{fa} then
imposes constraints on some of the coefficients of the
subleading terms in the expansion \rf{xa}:
\bea\la{MVV}
f_{10} =\ha f \ (f_c-1+J )  \ , \ \ \ \ \ \ \
f_{32} = {\textstyle \frac{1}{16}}f \ [ f^3 - 2 f^2 (f_c-1  +J)-  16 f_{21}] \ , \ \ ...
\eea
where dots stand for
similar expressions  for $f_{31}, f_{30}, f_{55}, ..., f_{50},$ etc.\foot{Note that 
using this expression for $f_{10}$ 
and defining $\td f_c = e^{ - f_c/f}$ to put \rf{xa} 
into the form \rf{gh}  one finds that 
$f'_{10} = \ha f \ (-1+J )$  (which, at strong coupling,  is subleading  to 
$f_{11}$ term  unless $J \sim \sql$).}
Again, these equations relate functions $f_{mk}(\l)$ defined as   power series in $\l$.
For twist $J=2$ we get simply
\bea\la{VV}
f_{10} ={1 \ov 2}  f  (f_c+1)  \ , \ \ \ \ \ \ \
f_{32} = {\frac{1}{16}}f \ [ f^3 - 2 f^2\ (f_c+1)-  16f_{21}] \ , \ \ ... \ .
\eea
These so called  MVV ~\cite{moch} relations were first observed
for twist 2  QCD   anomalous dimensions up to 3 loops.
The large $S$ expansions
for the known twist 2 and twist 3  SYM anomalous dimensions that we will
present  in  Appendix F
 are indeed
consistent with these relations, i.e.  with the reciprocity property of the function $\PP$.
\footnote{Three-loop tests of reciprocity for QCD and for the universal
 twist 2 supermultiplet in $\mathcal{N}=4$ SYM were discussed
 in~\cite{bk,dok2}. A four-loop test for
the   twist 3  anomalous dimension in the $sl(2)$ sector
was performed   in~\cite{bdm}. The case of twist 3
gauge field strength  operators was  analyzed in~\cite{Beccaria:2007pb} (at three loops)
 and in ~\cite{bf} (at four loops). In the latter
paper it was also proved that even the wrapping-affected
four loop result for the twist two operators ~\cite{Kotikov:2007cy}
is reciprocity respecting.}

\bigskip

It is natural to  expect that the functional relation  and the reciprocity property
should  hold  also  at  higher orders in small $\lambda$  expansion. Since the planar perturbation theory
should be convergent, they should then
 also  be visible at strong coupling  \ci{bk}, i.e.  in the large spin expansion
of the corresponding semiclassical string energies.

One may  also wonder   if  the reciprocity property
may   apply to  higher twist operators  above  the lower bound of  the band  \ci{bgk,bkp}.
If that were the case, it  could then  be  checked  also
at strong coupling on the example of the spiky string solution of \ci{kru}.

\

The agreement in  the {\it structure} of the large $S$ expansion found
in perturbative gauge theory and in perturbative string theory is already quite  remarkable.
 This agreement is non-trivial  since, as was already mentioned,
  the gauge-theory and string-theory
perturbative expansions are organized differently:
the gauge-theory limit is to expand in small $\l$ at fixed $S$
 and  then expand the $\l^n$ coefficients in large $S$,
while the semiclassical string-theory limit is to expand in large $\l$  with fixed $\S =
{ S \ov \sql}$  and then expand the $ 1 \ov (\sql)^n$  terms in $E$  in large
$\S$.
Even assuming these limits commute (which so far appears to be verified only
 for  the leading universal  $\ln S$ term)
the reason for the  validity of the
 functional relation \rf{fa}  and, moreover, of  the
 reciprocity property  \rf{rec} is 
  obscure on the semiclassical string theory side.

The functional relation \rf{fa} for the  anomalous dimensions  of Wilson-type operators
on the gauge theory  side was argued \ci{bk} to follow from the invariance  under the
 collinear $SL(2, R)$  subgroup of the conformal $SO(2,4)$ group.
 Given that this argument  is  based on the
  conformal symmetry, one may think  that it should
  then   apply also  on the string theory side.
  However, as we will review in Appendix A,
 the  realization of the conformal group  on
 states represented by classical spinning string
 solutions  in  global $AdS_5$  coordinates  is a priori
  different from the one used  on the gauge-theory side (which is 
  based on the collinear subgroup),  so that   the  direct connection is not obvious.
  The reason for  the reciprocity property on the string theory side
  is even far  less clear.

 If one   identifies  the energy $E$   and the spin $S$
of a string rotating in a plane  in global $AdS_5$
with dimension and Lorentz spin  of the  gauge theory
operator like $\tr ( \DD_+^S \Phi^J)$,   the functional relation
\rf{fa}  would then   imply   that  $\g =E-S-J$  should be a
 function of  $ s=\ha( E+ S)$, i.e.
\be  E-S-J = \PP ( E+S, J)   \ . \la{faw} \ee
As we shall discuss below (extending earlier observations in \ci{pt,bk}),
not only the structure of the large
 spin expansion on the string theory side happens to be   the same as on the gauge theory side
 but also
  its coefficients are indeed
  consistent  with the functional relation and the reciprocity
  for the minimal dimension case represented by the folded spinning string.
  This  will be demonstrated at the classical as well as 1-loop string theory  level.

  We shall  also show that  the functional relation but not the reciprocity
  appears to apply also  to  the case of the classical spiky string solution.

\bigskip

The rest of this  paper is organized as follows.

In section 2 we shall first  consider the large spin expansion of the classical energy
of folded spinning
string in $AdS_5$  and show that
the large spin expansion has the structure \rf{exx} and the functional and reciprocity
relations between the coefficients
are satisfied. We shall then include (in section 2.2) the  dependence on
the angular momentum $J$ in $S^5$ in the
``long string''  limit ($\J \ll \S$). In section 2.3
we shall study  the same large spin expansion for a spiky
string in $AdS_5$; in this case we shall find that
 the reciprocity condition is violated  which
should be  related to the fact that the
 corresponding operator has higher than minimal dimension
for a given spin.

In section 3 we shall return to the case of the folded spinning
string in $AdS_5$ (i.e. assume that $J$ is  negligible compared to $S$)
  and  compute the 1-loop correction to the energy expanded in
 large $S$, determining corrections to several  leading coefficients.
   As result,  we shall verify
that the string 1-loop corrections preserve the structure
 \rf{xax} of the large spin expansion and,
moreover,  that the reciprocity condition is satisfied beyond
the  string tree level.

In Appendix A we shall make some comments  on relation between different
realizations of conformal group.
In Appendix B and C we shall review the folded spinning string solution  and 
discuss  long-string or large-spin expansions 
used in the 1-loop computation in section 3. 
In Appendix D we shall give  details of large spin
expansions for $(S,J)$  string  considered  in section 2.
In Appendix E we shall  discuss some 
 consequences of the functional relation and the reciprocity
at  strong coupling, pointing out a subtlety in the definition of the latter
in the semiclassical string expansion.
In Appendix F we shall summarize the known weak coupling planar SYM results
for the large spin expansion of twist 2 and 3 anomalous dimensions up to 4-loop order
in the `t Hooft coupling.

\renewcommand{\theequation}{2.\arabic{equation}}
 \setcounter{equation}{0}

\setcounter{equation}{0}

\section{Large spin expansion: classical string theory  }


\subsection{Folded spinning string  with $J=0$ }

We shall start  with a discussion of the limit when the $S^5$  momentum $J$
 of the string state can be ignored, i.e. we shall
 concentrate only on  the $AdS_5$ spin $S$ dependence of the string energy.
 This is the
 limit  when the twist of the gauge theory operator is
 sufficiently small compared to the Lorentz spin.

 We review
  the   folded spinning string solution \ci{gkp}  in Appendix B.
The integrals of motion
 are the energy $E=\sl\,\E$ and the spin
  $S=\sl\,\S$, which  can be expressed in terms of the elliptic functions
  $\EE$ and $\KK$ of an
   auxiliary variable $\eta$  \foot{Here we follow the notation of
    \ci{bfst}. Equivalently,  one can express the conserved charges in
     terms of the hypergeometric functions as in Appendix B.}
\bea\la{EmS}
\E-\S&=&\frac{2}{\pi}\sqrt{\frac{1+\eta}{\eta}}\left[\EE\left(-
\frac{1}{\eta}\right)\left(\frac{1}{\sqrt{1+\eta}}
-1\right)+\KK\left(-\frac{1}{\eta}\right)\right]\ , \\\label{pin}
\S&=&\frac{2}{\pi}\sqrt{\frac{1+\eta}{\eta}}\left[\EE\left(-\frac{1}
{\eta}\right)-\KK\left(-\frac{1}{\eta}\right)\right] \ .
\eea
To find the energy in terms of the spin one is to solve for  $\eta$.
Here we are interested in the large spin expansion which corresponds
to the long string limit (when the string   ends are close to
 the boundary of $AdS_5$).
For such long string  one has $\eta\to 0$.
  Solving (\ref{pin}) for small  $\eta$  and substituting it
  into (\ref{EmS}) one finds for $\E$  as a function of ${\cal S}$
\be
&&{\cal E}= \S+\frac{\ln\bar \S-1}{\pi} + \frac{\ln\bar \S-1}{2\,\pi^2\,\S}-\frac{2\ln^2\bar
 \S-9\ln\bar \S+5}{16\,\pi^3\S^2} \no\\
&& \ \ \ \ \ \ \ \ \ \ \
+\ \frac{2\ln^3\bar\S-18\ln^2\bar\S+33\ln\bar\S-14}{48\,\pi^4\,\S^3}+...\ , \ \ \ \ \ \  \ \ \
\bar \S\equiv 8\,\pi\,\S \ ,
\label{EmS1}\ee
as was already claimed  in  \rf{exx}.

The functional  relation \rf{fa},\rf{faw}  implies that $\E-\S$   should be a function
of $\E+\S$.   It is not immediately obvious
from \rf{pin} (or from the form of the exact solution in global $AdS_5$ coordinates)
 why   such a relation should be natural for any value of $\S$.
 Still,  the coefficients of the leading $ ({\ln \S \ov \S})^m $
 terms in \rf{EmS1} happen, indeed, to be  consistent
  with such a relation,
    with the leading term in the function
    $\PP$ being simply the logarithm
  (cf. \rf{lea})
\be\label{lead}
E-S=
 \frac{\sqrt{\l}}{\pi}\ln\Big[S+\frac{1}{2}\frac{\sqrt{\l}}{\,\pi}\ln  S +...\Big]+... \ .
\ee
Furthermore, it is possible
 to verify that  the expansion of $E-S$  also satisfies the
 reciprocity property \rf{rec},\rf{VV}.
The
large $\S$ expansion of the function $\PP$ (its leading term in the strong-coupling limit)
 is much simpler
than that of   the anomalous dimension $E-S$ in  (\ref{EmS})
and contains only {\it even} powers of  $C^{-1} \sim \S^{-1}$ (see \rf{casimir})
\be\label{P01} \PP = \sql\ \tP \ , \ \ \ \ \ \ \ \
{\tP}(\S)={1 \ov \pi} \Big[ \ln \bar \S-1
+\frac{\ln \bar \S+1}{16 \pi ^2\S^2 }  + {\cal O}\left({1\ov \S^4}\right)\Big] +  {\cal O}({1 \ov \sql})  \ .
\ee
Equivalently, we find that  the MVV-like  relations \rf{VV} are
satisfied.\footnote{The definition of reciprocity condition
in string semiclassical expansion  is discussed in Appendix E.}

A more systematic analysis  of the reciprocity  (parity invariance)
 property  of the function $\PP$ is possible with the  help of   an integral
 representation for it.
Using that \rf{fa} implies
$\tP(\S')= \tg\left(\S'-\textstyle{\frac{1}{2}}\tP(\S')\right)
$, where $\S'= S + \ha \tg (\S)$,   $\tg(\S) = \E - \S$,
and renaming $\S' \to \S$
we have
\be\label{LB}
\tP(\S)=\frac{1}{2\pi \,i}\oint_{\Gamma} d\o \ \tg(\o)\ \frac{1+\ha {\tg'(\o)}}{
\omega-\S+\ha {\tg(\o)}}\ ,
\ee
where the contour $\Gamma$ encircles
 the pole of the integrand and prime stands for derivative.\footnote{The expression that
  multiplies $\tg$ in the integrand has residue $1$, so that
   the integral is $\tg$ evaluated at the
   pole $\o=\S-\ha {\tg}$.
   Then defining $x=\S-\frac{1}{2}\tP(\S)$
    we have   $2 \S-2x=\tg$ which coincides with the
     equation for the pole with $x=\omega$.
     Note that assuming $\PP$ exists,
one can  formally reconstruct  it
from
$\g$ using \ci{bk}
${\PP}(S)=\sum_{k=1}^\infty \frac{1}{k!}\left(\textstyle{-\frac{1}{2}}\partial_S\right)^{k-1}[\gamma(S)]^k
= \gamma-\frac{1}{4}\,(\gamma^2)'+\frac{1}{24}\,(\gamma^3)''
+ \cdots . $
    This relation also  arises
     by expanding the denominator in \rf{LB} in small $\tg$ and integrating
     the resulting   series.}
It is  natural to replace  the variable $\omega$ in (\ref{LB}) with
the expression (\ref{pin}) for the semiclassical spin $\S(\eta)$
\be\label{B2}
\tP(\S)=\frac{1}{2\pi \,i}\oint_\Gamma\,d\eta\, \tg(\eta)\,\frac{{\tilde s}'(\eta)}{{\tilde s}(\eta)-\S} \ ,
\ee
where ${\tilde s}(\eta)\equiv\S(\eta)+\ha \tg(\eta)= \ha (\E + \S)$ is
the ``conformal spin'' expressed  in terms of the
 semiclassical quantities. The  integral then gives
 the function $\tg$ evaluated at  the zero of the denominator;
 this  is the same as the  statement that the anomalous dimension as a function of
  the Lorentz spin is, effectively, a function of
  the conformal spin ${\tilde s}$.

To verify the reciprocity property  of the function $\tP(\S)$
 in  (\ref{B2}) it is useful to redefine  the variable $\eta$
 as\footnote{This choice is not unique.
 An analogous  transformation was used in~\cite{bk}.}
   $\eta\rightarrow -1+16\eta+\sqrt{1+256\,\eta^2}$
   and examine the large $\S$ or small $\eta$ limit of the expressions.
One finds  that  $\tg(\eta)$  is a series in even powers of $\eta$
\be\label{ven}
\tg(\eta)=-\frac{1+\ln\eta}{\pi}+\frac{4(\ln\eta+12)}{\pi}\eta^2-\frac{6(62\ln\eta+777)}{\pi}\eta^4+... \ ,
\ee
while the expression for the conformal spin runs in odd powers of $\eta$
\be\label{nodd}
{\tilde s}(\eta)=\frac{1}{8\pi\eta}+\frac{11+2\ln\eta}{2\pi}\eta -\frac{877+92\ln\eta}{2\pi}\eta^3 +... \ .
\ee
From the equation for the pole of the integrand in \rf{B2},
 ${\tilde s}-\S=0$, one can find  the parameter $\eta$ in terms of
 the spin $\S$,  concluding that it is given by a power series in  odd negative powers of $ \S$.
As a result,  $\tP(\S)$, which is  same as $\tg(\eta)$ evaluated at  the pole, should
 also run only in even negative powers of $\S$ or $\C = { C \ov \sql}$ (cf. \rf{casimir}).

The above discussion has a straightforward   generalization to the multifolded spinning string case.
The leading terms in the  large $\S$   expansion of the energy of a string with $m$ folds
are (see Appendix D)
\be\label{multifoldnu0}
\E-\S= \frac{m}{\pi}\,\Big[\ln\tS-1+\frac{4}{\tS}\,(\ln\tS-1)-\frac{4}{\tS^2}\,(2\ln^2
\tS-9\ln \tS+5)+...\Big], \ \ \ \ \ \ \   \tS\equiv { 8 \pi \ov m} \S \ .
\ee
 In this case
it is possible to show again
 that  the large $\S$  expansion  is  consistent   with the
 reciprocity property.


\subsection{Folded spinning  string  with $J\not=0$}

Let us now consider the case when the $S^5$ angular momentum of the string is not negligible
 compared
to $S$,
i.e. when the string state  is dual to an operator
 with large spin $S$    and large twist $J$.
The corresponding  charges are the energy $E=\sl \,\E$ and the two
 angular momenta $S=\sl\,\S$ and $J=\sl\, \n$ \ \ci{ft1,bfst}:
\be\label{foldinform1}
&&\ \ \ \ \ \ \ \ \  \ \ \  \E=\k+\frac{\k}{\omega}\S\ ,~~~~~~~~~~~~~
\frac{\o^2-\J^2}{\k^2-\J^2}\equiv1+\eta\ ,
\\
\label{foldinform2}
&&\sqrt{\k^2-\J^2}=\frac{2}{\pi\,\sqrt{\eta}}\KK\left(-\textstyle{\frac{1}{\eta}}\right),
~~~~~~~~~~~~~~~~
\S=\frac{2\pi\,\sqrt{\eta} \omega}{\sqrt{\k^2-\J^2}}\,\left[\EE\left(-
\textstyle{\frac{1}{\eta}}\right)-
\KK\left(-\textstyle{\frac{1}{\eta}}\right)\right]
\ee
Here $\k$ and $\omega$ (or $\eta$)  are parameters of the classical solution which
 should we eliminated to find $\E$ as a function of $\S$ and $\J$.

We will be  interested in large $\S$  expansion   with   $\S \gg \J$
since only  in this case the expansions  like \rf{xa},\rf{rec}, i.e. going
  in the  inverse powers of $\S$ with the
coefficients being polynomials  in $\ln \S$,  will apply (see also \ci{ft1,bk}).

In the large $\S\gg \J$ or  \emph{long string} limit, when  $\eta\ll 1$,
one should distinguish between  ``small'' or  ``large''  $\J$ cases \ci{ft1,ftt}.
In the ``slow long string" approximation  (corresponding to taking $\S $ to be large  with
$\ell \equiv { \J \ov \ln \S} $ fixed and then expanding in powers of $\ell$)
the leading  terms in  the semiclassical energy read (cf. \rf{EmS1})
\bea\label{enslow}
\E-\S-\J&\approx&\frac{1}{\pi} (\ln{\tS}-1)
+\frac{\pi\,\J^2}{2\,\ln\tS}
-\frac{\pi^3\,\J^4}{8\,\ln^3\tS}\big(1-\frac{1}{\,\ln\tS}\big)+...\\\nonumber
&&~~~~+\frac{4 }{\tS}\Big[{ 1 \ov \pi} (\ln{\tS}-1)
+\frac{\pi\,\J^2}{2\ln^2{\tS}}
-\frac{3\pi^3\,\J^4}{4\ln^4{\tS}}\big(1-\frac{2}{3\,\ln\tS}\big)+...\Big]\\\nonumber
&&~~~~-\frac{4}{\tS^2}\Big[{1\ov \pi} (2\ln^2{\tS}-9\ln{\tS}+5)+\pi\,\J^2
\Big(1+\frac{3}{2\ln{\tS}}
-\frac{1}{\ln^2{\tS}}-\frac{2}{\ln^3{\tS}}\Big)+...\Big]
\eea
where  $\tS\equiv 8\pi\S$,  and dots stand  for higher order
  corrections depending on  $\J$. \foot{Note that  the  leading terms  in expression of the previous subsection \rf{EmS1}
 dominate in the limit when
 $\frac{\J^2}{\ln\S} \ll { \ln \S \ov \S }$.}

In the case of ``fast long string'', when
 $\ln\S\ll\n\ll \S$, the corrections to the energy read
\bea\nonumber
&&\!\!\!\!\!\!\!\!\!
\E-\S-\J\approx\frac{1}{\pi^2\,\J}\Big[{ 1\ov 2} {\ln^2\hS}-\ln\hS+\frac{4\ln\hS}{\hS}+
\frac{4}{\hS^2}\big(-2\ln\hS+1+\frac{3}{\ln\hS}+\frac{2}{\ln^2\hS}+...\big)+...\Big]\\\label{enfast}
&&
~~~~~~~~~~+\frac{1}{\pi^4\J^3}\Big[-\frac{\ln^4\hS}{8}-\frac{2}{\hS}\big(3\ln^2\hS+
\ln\hS+1+\frac{1}{\ln\hS}+\frac{1}{\ln^2\hS}+...\big)\\\nonumber
&&~~~~~~~~~~~~~~~~~~~~~~~ -\frac{2}{\hS^2}\big(2\ln^3\hS-19\ln^2\hS+11\ln\hS+
13+\frac{13}{\ln\hS}+\frac{11}{\ln^2\hS}+...\big)+...\Big]
\eea
where $\hS\equiv
{8S\ov J} ={8\S\ov \J}\gg 1$. Dots in the square brackets
 indicate corrections in $1/\hS$, corrections in  $1/\ln\hS$ can
  be added in the round brackets and  terms like $ \ln(\ln\hS)$
  have been neglected.

The leading terms  here can be summed up as  \ci{bgk}
\be
\E- \S = \sqrt{ \J ^2 + {1 \ov \pi^2} \ln^2 { 8 \S \ov \J }}  + ...\ ,  \ee
where ${  \ln \S \ov \J } \ll 1 $ plays the role of  an expansion parameter.

Notice that in contrast to the  slow long string case where  the expansion
\rf{enslow} has the same structure  as in \rf{qw}, in the fast long string
case \rf{enfast} we get  higher powers of $\ln \S$  not suppressed by $\S$,
and so this case (cf. also  its discussion  in Appendix D)
is somewhat outside our main  theme here.

To study  the properties of the subleading corrections,
 one may  again make  use of the integral representation for the
 functional relation as in \rf{LB}.
The discussion will  apply to  both the  ``slow'' and the ``fast'' long string limits.
Here the  ``conformal spin''
is  ${\tilde s}=\ha ( \S + \E)=\S+\ha \n+\ha \tg, $
while the ``semiclassical''
value of the
 Casimir operator in \rf{casimir} is
 $\C= {C \ov \sql}  \approx \S + \ha \J$.
Then
 the integral in (\ref{B2}) can be written as
\be\label{LB2}
\tP(\C)=\frac{1}{2\pi \,i}\oint_\Gamma\,d\eta\, \ \tg(\eta)\,\frac{{\tilde s}'(\eta)}
{{\tilde s}(\eta)-\C}\ , \ \ \ \ \ \ \ \ {\tilde s}(\eta)=\S(\eta)+\ha \tg(\eta)   \ .
\ee
 After a redefinition  of $\eta$ one can then show   that
 the  expansion of $\PP$ in large $\C$
  runs only in even negative powers of $\C$.
Some details are given  in Appendix D.
In the  kinematic region of ``fast'' long strings,  with
$1 \ll \ln\S\ll\n\ll\S$,  this parity invariance  property
was already demonstrated in a closely related way
 in~\cite{bk}.



\subsection{Large spin expansion of  energy of a spiky string in $AdS_5$}

Let us now  consider the spiky spinning string  in $AdS_5$ \cite{kru}, and
 find  corrections to the leading $\ln S$ term in its large spin expansion.

The integrals of motion here are  the  difference between the position
of the spike and of the middle of the
valley between the two spikes, the
 spin and the energy~\cite{kru}~\footnote{In the case of
the multiply folded string with $n$ spikes
  multiplying formulas
 one should multiply
  (\ref{Ener}) by the number  $m$ of the folds, and
  use that in this case  $\Delta\theta=\frac{\pi}{\,n\,m}$.
  As a result, one is simply  to substitute  $n\to n\,m$.}
\bea
\label{Th}
 \Delta \theta ={ \pi \ov n}=  \frac{\sinh 2\rho_0}{\sqrt{2}\sinh\rho_1}\frac{1}{\sqrt{u_1+u_0}}
 \left[\Pi(\frac{\pi}{2},\frac{u_1-u_0}{u_1-1},p)-\Pi(\frac{\pi}{2},\frac{u_1-u_0}{u_1+1},p) \right]\ ,
 \eea
\be \label{Spinsp}
\S =  \frac{n\ \cosh\rho_1}{\sqrt{2}\ \pi \  \sqrt{u_1+u_0}}\ \bigg[
    -(1+u_0) {\KK}(p) + (u_1+u_0) {\EE}(p) -
  \frac{u^2_0-1}{u_1+1} \Pi(\frac{\pi}{2},\frac{u_1-u_0}{u_1+1},p) \bigg] \ ,\la{sq} \ee
\bea\label{Ener}
\E-\omega \S =  \frac{n \sqrt{u_1+u_0}}{\sqrt{2}\ \pi \ \sinh\rho_1}
  \ \left[ {\KK}(p)-{\EE}(p)\right] \ ,
\eea
where $n$ is the number of the spikes and
\be
u_0=\cosh 2\r_0\ ,~~~~~~~~u_1=\cosh2\r_1\ ,
~~~~~~~~\o=\coth\rho_1\ ,
~~~~~~~~p=\sqrt{\frac{u_1-u_0}{u_1+u_0}} \ .
\ee
The string is rigidly rotating with
 the  radial coordinate being
 $\rho=\rho(\sigma)$, with  $\rho_0$ and $\rho_1$ as its
 minimal and maximal  values (positions of the  bottom of the valley between the spikes and
 the spikes themselves).  $\rho_0$ and $\rho_1$ are related by the condition \rf{Th}. Solving
 for the remaining free parameter gives $\E=\E(\S,n)$.

 The large  spin limit corresponds to  $\r_1\to\infty$, {i.e.} to the case when the ends
of the spikes approach the boundary of $AdS_5$.
Let us  set
\be
y=e^{-2\r_1} \ ,
\ee
and expand in  $y\to 0$.  Then, at leading order,  $\Delta
 \theta=\frac{\pi}{n}= \arcsin\frac{1}{u_0}+{\cal O}(y)$
  implies $u_0=\cosh 2 \rho_0 = \csc\frac{\pi}{n}$
and
\bea
\E-\S=-\frac{n}{2\pi}\ln y+{\cal O}(y)\ , \ \ \ \ \ \
\S=\frac{n }{4\pi}\frac{1}{y}+{\cal O}(\ln y)\ ,
\eea
i.e.
\be\la{spil}
\E-\S=\frac{n}{2\,\pi}\ln\frac{16\,\pi\, \S}{n} + ... \ .
\ee
This is the result already found in~\cite{kru},  which  reduces to the case of the
 folded string when  $n=2$.~\footnote{For $n=2$  we have
  $\Delta\th=\frac{\pi}{2}$ (i.e. the angle between spikes is $\pi$), and thus
   $\rho_0=0$ or $u_0=1$.}
Expanding further near   $y\simeq 0$ one gets
\bea
\S&=&\frac{n  }{4\,\pi}\Big(\frac{1}{y}+\ln y+1-2 \sqrt{u_0^2-1} \arccos
 \sqrt{\frac{u_0+1}{2\,u_0}}+\ln\frac{u_0}{4}\Big)+ ...\ , \\
\Delta\th&=&\frac{\pi}{n}\ = \arcsin\frac{1}{u_0}+y \Big(2 \arcsin\sqrt{\frac{u_0+1}{2\,u_0}}
-\pi \Big) + ...\ ,
\eea
where the second equation can be used to fix $u_0$ in terms of $y$  and the number of spikes $n$. Eliminating then $y$ in favor of $\S$,
  we have from (\ref{Ener})
\bea\label{EmSspikes}
&&\E-\S=\frac{n}{2\pi}\Big[ \ln\tS  + p_1   
+\frac{4}{\tS}\left(\ln\tS + p_2 \right) -\frac{4}{\tS^2}\left(2\ln^2\tS  + p_3 \ln\tS+ p_4  \right) \no \\
&&\ \ \ \ \ \ \ \ \ \ \ \ \ \ \ \ \ \ \ \
+\ \frac{32}{3\,\tS^3}\left(2\ln^3\tS  + p_5  \ln^2\tS+ p_6  \ln\tS  + p_7   \right) +...\Big]\ ,
\eea
where   
\be\la{opr}
&& \ \ \ \ \ \ \ \ \ \ \ \ \ \ \ \ \ \ \ \ \ \ \ 
\tS={16\,\pi\ov n}\S  \\ && p_1  = -1+\ln\sin\frac{\pi}{n} \ ,~~~~~~~~~~~~~~~~~
p_2= -1+\ln\sin\frac{\pi}{n}+\frac{\pi(n-2)}{2n}\cot\frac{\pi}{n}\ ,\\
&&p_3= - 10 +  \frac{2\pi (n-2)}{n} \cot \frac{\pi }{n}  - 2  \cot^2\frac{\pi }{n}  
 -4  \ln \csc \frac{\pi }{n}   +  \csc^2\frac{\pi}{n} ,
\no \\
&& p_4=  6 - \csc ^2\frac{\pi}{n}+ \frac{\pi ^2 (n-2)^2}{2 n^2} -\frac{4\pi (n-2)}{n} \cot \frac{\pi }{n}  + 
\cot^2\frac{\pi }{n} \Big[\frac{\pi ^2 (n-2)^2}{n^2} +1\Big] \no \\
&&\ \ \ \ \ \ \ \ + \
\ln \csc \frac{\pi }{n}\ \Big[2 \cot^2\frac{\pi }{n}- \frac{2 \pi(n-2)}{n}\cot\frac{\pi }{n}
- \csc ^2\frac{\pi }{n} +2 \ln \csc \frac{\pi}{n}+10\Big]
\ ,\\
&& p_5= -18 + \O(n-2) \ , \ \ \ \ \
p_6= 33 + \O(n-2) \ , \ \ \ \ \
p_7=  -14 + \O(n-2) \ . 
\ee
It is easy to check  that (\ref{EmSspikes}) coincides with the energy (\ref{EmS1}) for the folded string
 in $AdS_5$   when $n=2$.\foot{Note also that the form of $p_1$ is consistent with the interpretation 
 of the subleading
 term in the energy in \ci{dor}.}

Retaining in (\ref{EmSspikes}) only the dominant contributions at  each order of
 the above expansion we obtain
\bea
\E-\S= \frac{\,n}{2\pi}\ln\S  +\frac{n^2}{8\,\pi^2\S} \ln\S
-\frac{n^3}{64\,\pi^3\S^2} \ln^2{\S}    +\frac{n^4}
{384\,\pi^4\S^3}  \ln^3\S +...\ .
\eea
This  may be rewritten as
\be\la{spk}
E-S= \frac{\sqrt{\l}\,n}{2\pi}\ln\Big[S +\frac{1}{2}\frac{\sqrt{\l}\,n}{2\,\pi}
\ln S \Big]+... \ ,
\ee
implying that the functional relation is satisfied (cf. \rf{lead}).

However, the reciprocity property  is not respected in this case. Indeed,
 the analog of the function ${\tP}(\S)$ in \rf{P01}  has the following expansion
\bea\label{Pspi}
\tP(\S)&=&\frac{n}{2\pi}\Big[\ln\tS  + q_1 
+ \frac{q_2}{\tS} +\frac{1}{\tS^2}(q_3\,\ln\tS+q_4)  +\frac{1}{\tS^3}(q_5\,\ln\tS+q_6) ...\Big]+... \ ,
\eea
where 
\bea
  q_1&=&-1+\ln\sin\frac{\pi}{n}\ , \ \ \ \ \ \ \ q_2= \frac{2 \pi (n-2)}{n} \cot\frac{\pi}{n}\ ,\ \
  \   \ \ \ 
  q_3=4\csc^2\frac{\pi}{n}\ ,\\  
  q_4&=&4+2 \pi^2\,\big(\frac{n-2}{n}\big)^2(1-2\csc^2\frac {\pi}{n})+ 4\ln\sin\frac{\pi}{n}\csc^2\frac{\pi}{n}\ ,\\
   q_5&=& \O(n-2)\ , \ \ \ \ \ \ \ \ \ \ \ \ q_6= \O(n-2)\ , 
   \la{suu}
\eea
where $q_5,q_6$ are non-zero for $n \not=2$.
The expansion (\ref{Pspi}),  even if considerably simpler 
compared  to the energy (\ref{EmSspikes}), is   not parity invariant under $\S\to-\S$.
It is interesting though that higher powers of  $\ln \S$ appear 
to cancel in the subleading terms in \rf{Pspi}.\foot{This  feature of the $\tP$-function is
 in a marked contrast with 
 the anomalous dimension, whose large $S$ expansion includes growing powers of $\ln S$
 in the coefficients of $1/S^n$ terms.
 This reduction of singularity of the large $S$ expansion 
 of $\tP$  was observed also at weak coupling~\ci{bdm,bf,bkp}.}
  The parity invariance is restored  in the case of the folded
   string when $n=2$, where indeed (\ref{Pspi}) coincides with (\ref{P01}).

This breakdown of parity invariance for a string with  $n > 2$ spikes
is not totally surprising, as such spiky string  should  correspond  to
an operator with \emph{non-minimal} anomalous dimension for a given spin,
 while the reciprocity  was checked at weak coupling only for the minimal
 anomalous dimensions.
Indeed, anomalous dimensions of operators
 of twist higher than two with trajectories close to the
 upper boundary of the band also do not respect the reciprocity
 as was  seen recently in the  twist three case at weak coupling in
 \cite{bkp}.
 
 It is interesting that our strong-coupling result \rf{spk},\rf{Pspi} has close similary with 
 weak-coupling one found for $n=3$  in \ci{bkp}: the functional 
  relation \rf{spk} is still satisfied, 
 and the parity invariance is broken at level $1/S$.
 Interestingly, the $1/S$ coefficient $\sim  n q_2$ in \rf{Pspi} (cf. \rf{opr}) 
 is proportional,  for $n=3$, to $\sqrt 3$, which is, suprisingly,  
 the same  factor appearing also 
in the corresponding expression at  weak coupling \ci{bkp}.\foot{We thank G.
 Korchemsky for this observation.} 
 In general, this  coefficient should be a function of $\l$
 interpolating from weak to strong coupling 
  but its   dependence on  $n$
 might be the same  for any $\l$.

\renewcommand{\theequation}{3.\arabic{equation}}
 \setcounter{equation}{0}

\section{Large spin expansion of  folded string energy: 1-loop order}
\label{sec:1loop}

Let us  now go back to the folded spinning string case of section 2.1   and
compute the leading $1$-loop corrections to its  energy \rf{EmS1}
 in the large $\mathcal{S}$ expansion. We shall follow the  general approach
 for computation of  quantum string  corrections  developed in \ci{ft1}
 where the 1-loop shift of the  $\ln \S$ term  was  found.\foot{The
 2-loop correction
 to the scaling function  was found in
 \ci{rtt,rt}; a generalization to non-zero $J$ was considered in
 \ci{ftt,rt2} (parallel  results from the  Bethe ansatz were  found  in \ci{kri,grm}).}
 We shall find the 1-loop corrections to the subleading terms in \rf{EmS1} by
 applying a  perturbative procedure similar to the one used
 in  \cite{tt} in the small spin expansion case.

 Our aim will be to verify that:\ (i) the structure of the large spin expansion \rf{EmS1}
 remains  the same also with the 1-loop corrections included,  and\  (ii) the
 constraints on the coefficients
 imposed by the functional relation and the reciprocity  remain to be satisfied
 at the 1-loop order.

The fluctuation action
in the  conformal gauge  expanded to
quadratic order in fluctuations near the folded spinning string
solution
$
\bar S=-\frac{\sqrt{\lambda}}{4 \pi}\int d \tau \int_0^{2 \pi} d \sigma\ \bar {L}
$  has  the following 
bosonic part (see \ci{ft1} and Appendix B)
\begin{eqnarray}
\bar {L}_B=&-& \partial_a \td {t} \partial^a \td {t}- \mu_t^2 \td {t}^2 +   \partial_a \td {\phi}
 \partial^a \td {\phi}+ \mu_{\phi}^2 \td {\phi}^2\nonumber\\
&+& 4 \td {\rho} (\kappa \sinh \rho\ \partial_0 \bar {t} - w \cosh \rho\ \partial_0 \bar {\phi})+
\partial_a \td {\rho} \partial^a \td {\rho}+\mu_{\rho}^2 \td {\rho}^2\nonumber\\
&+& \partial_a {\beta}_u \partial^a {\beta}_u +\mu_{\beta}^2 {\beta}_u^2 +
 \partial_a {\varphi} \partial^a {\varphi}+\partial_a {\zeta}_s \partial^a {\zeta}_s \ ,   \label{lag}
\end{eqnarray}
where
\begin{equation}
\mu_t^2= 2 \rho'^2 -\kappa^2, \ \ \quad \mu^2_{\phi}=2 \rho'^2 -w^2, \ \
\quad \mu^2_{\rho}=2 \rho'^2 -w^2-\kappa^2,\ \
\quad \mu_{\beta}^2=2 \rho'^2  .    \label{as}
\end{equation}
Here $\beta_u$ ($u=1,2$) are the two $AdS_5$ fluctuations
transverse to the $AdS_3$ subspace in which the string is moving, while
 $\varphi,\zeta_s$ ($s=1,2,3,4$) are fluctuations in $S^5$.
 The fermionic part of the quadratic fluctuation Lagrangian   can be put into the form \ci{ft1}
 \be \label{feq}
 \tilde{L}_{_F}  =  2 i ( \bar  \Psi \gamma^a \partial_a \Psi - \mu_{_F} \bar  \Psi \Gamma_{234} \Psi) \ ,
 \ \ \ \  \ \ \ \ \ \
 \mu_{_F}= \rho'  \   , \ee
 and  can be interpreted as describing a system of 4+4   2d Majorana  fermions with
  $\sigma$-dependent mass
 $ \rho' $.
 As explained in \ci{tt}, after squaring the corresponding ``Dirac'' operator, 
  the fermionic contribution to the 1-loop 
 partition function can be represented as 
 \be   - \ha \Big(  4 \ln \det \Delta_{\F+} +  4 \ln \det \Delta_{\F-} \Big) \ , \ \ \ \ \ 
 \ \ \ 
 \Delta_{\F\pm}  \equiv  - \partial^a \partial_a     \pm \r'' + \r'^2  \ . 
 \la{fero}
\ee 
In the leading-order  computation in the long-string limit that we are going to discuss below 
the  term $ \pm \r''$ in the effective  fermionic mass squared term in \rf{fero} 
can be ignored:  as follows from \rf{roo}, 
$\r''= O(\eta)$  and  since according to \rf{fero}
half of fermions   has $+\r''$ and half  $-\r''$ in their 
mass term, the leading  $O(\eta)$  contribution to the partition function 
can come only from the  $\r'^2$ term in $\Delta_{\F\pm}$.
We shall assume this  when  writing the fermionic 
contribution below.

 Switching  to euclidean signature ($\tau \rightarrow i \tau$),
 the $1$-loop correction to the energy can be  found from the  2d  effective action
\be \la{loi}
E_1 = {\Gamma_1 \ov \kappa \T}  \ , \ \ \ \ \ \   \qquad \T \equiv \int d \tau \to \infty   \ . \ee
Since the spinning string solution is stationary,  both
the bosonic and the fermionic fluctuation Lagrangians do not depend on
$\tau$;  thus, as in \cite{tt},  we may compute the relevant $2d$ functional
determinants by reducing
 them to $1d$ functional determinants using 
\begin{equation}
\det[-\partial_1^2 - \partial_0^2 + m^2]=
\T
\int {d{\omega}\ov 2\pi} \  \det [-\partial_1^2+\omega^2+m^2] \ ,\la{kou}
\end{equation}
where $m^2$  is  a generic mass term which may depend  on $\sigma$.

Given that  $\rho(\s)$ is a complicated function (see \rf{sol}),
 we are unable to determine the fluctuation spectrum exactly,
and,  as in \ci{tt},  we will resort to  perturbation theory in $1 \ov \S$ or in
parameter $\eta$ determining the maximal string length (see Appendix B and C).
In \rf{loi} we have  from  \rf{defa}
\be
\kappa= \ka  - \frac{\eta}{4 \pi}( \pi \ka   -2)+\O(\eta^2) \ , \ \ \ \ \ \ \
\ka \equiv  \frac{1}{\pi}\ln \frac{16}{\eta} \ .
\label{appa}
\ee
 $\G_1$ will  also be expected to have expansion
 in powers of $\eta \sim {1 \ov \S}$ (see \rf{spin})
with the coefficients containg   powers of $\ln \eta$.

To proceed,  we need to expand the fluctuation Lagrangian in small
 $\eta$ corresponding to large $\S$. Some relations
needed below can be found  in Appendix B. Let us first
perform (as in \cite{ftt}) the following rotation $(\bar t, \bar \phi) \to (\xi,\chi)$:
\begin{equation}
\xi =- \tilde{t}\ \sinh \rho + \tilde{\phi}\ \cosh  \rho,\
\qquad \chi=-\tilde{\phi}\ \sinh  \rho   +
  \tilde{t}\ \cosh \rho   \  .
\end{equation}
Then  the fluctuation Lagrangian takes the form 
\begin{eqnarray}
&&\!\!\!\!\!\!\!\! \bar {L}_B=- \partial_a {\chi} \partial^a  {\chi}+ (\mu_{\phi}^2 \sinh^2 
\rho -\mu_{t}^2 \cosh^2 \rho+\rho'^2)  {\chi}^2 +   \partial_a  {\xi}
 \partial^a  {\xi}+ (\mu_{\phi}^2 \cosh^2 \rho - \mu_{t}^2 \sinh^2 \rho - \rho'^2) \xi^2\nonumber\\
&&\!\!\!\!+  4 \bar {\rho} (\kappa \sinh^2 \rho\  - w \cosh^2 \rho) \partial_0 \xi+
\partial_a \bar {\rho} \partial^a \bar {\rho}+\mu_{\rho}^2 \bar {\rho}^2 +2 \rho' 
(\chi\xi'- \xi\chi')+ \chi \xi (\mu_{\phi}^2-\mu_{t}^2)\sinh 2 \rho\nonumber\\
&&\!\!\!\! +2 \tilde{\rho}\dot{\chi}(\kappa- w) \sinh 2 \rho+ \partial_a {\beta}_u \partial^a 
{\beta}_u +\mu_{\beta}^2 {\beta}_u^2 +
 \partial_a {\varphi} \partial^a {\varphi}+\partial_a {\zeta}_s \partial^a {\zeta}_s    \label{lag1}
\end{eqnarray}
The reason for  this rotation is that in the subsequent
 small $\eta$ expansion  the bosonic fluctuation Lagrangian at order 
  $\O(\eta^0)$ will become $\sigma$-independent, i.e.  will  have constant 
  coefficients as at the leading order in long-string expansion considered in  \ci{ft1,ftt}.
\foot{As discussed below, we shall ignore the contribution of the turning points 
at $\s={\pi \ov 2}$ and $3\pi \ov 2$ and will treat the fluctuation problem 
separately on each ``quarter-string'' interval.}
 
  Expanding the solution for $\r(\s)$ and the parameters $\k$ and $w$ in  small $\eta$
  (see Appendix B), 
  the bosonic fluctuation Lagrangian becomes 
  $\tilde{L}_B=\tilde{L}_0 +
 \eta \tilde{L}_1+...,$ where
\begin{eqnarray}
\tilde{L}_0=&-&\partial_a \chi \partial^a \chi + \partial_a \xi \partial^a \xi +  2 \ka
 \ \chi \xi'-   \ 2 \ka  \ \chi' \xi- \ 4 \ka  \ \tilde{\rho} \dot{\xi}\nonumber\\
&+&  \partial_a \tilde{\rho} \partial^a \tilde{\rho} + \partial_a {\beta}_u \partial^a
{\beta}_u +2 \ka^2  {\beta}_u^2 +
 \partial_a {\varphi} \partial^a {\varphi}+\partial_a {\zeta}_s \partial^a {\zeta}_s \ ,   \label{lagr}
\end{eqnarray}
and
\begin{eqnarray}
\tilde{L}_1&=& -\ka^2  \cosh (2 \ka \sigma) \xi^2
- \ka^2 \cosh(2 \ka \s) \tilde{\rho}^2-   \ka^2 \sinh(2 \ka \s)\ \xi\chi\
 - \frac{\ka}{\pi}[\ka \pi \cosh(2 \ka \s)- 2] {\beta}_u^2\nonumber\\
&+& (\chi \xi'- \xi \chi') [\frac{1}{\pi}-\frac{\ka}{2}\cosh (2 \ka \sigma)]- \tilde{\rho}
 \dot{\chi}\ka \sinh (2 \ka \sigma)-\tilde{\rho}\dot{\xi}[\frac{2}{\pi}+\ka \cosh (2 \ka \sigma)] \ .  \label{nxt}
\end{eqnarray}
As already mentioned,  the 1-loop
 effective action can be expressed in terms of
   1d functional determinants (with  $\partial_0
  \rightarrow i \omega$, see \rf{kou}).
  We shall denote the quadratic fluctuation operator in the 
  coupled $(\chi, \xi, \td \rho)$ sector as  $Q_{\omega}$.
 Since  $\tilde{L}_0$ has  constant coefficients,
 the leading part of the 
  fluctuation operator coming from 
  $\tilde{L}_0$  can be written as 
\begin{eqnarray}
Q^{(0)}_{\omega}=\left(
               \begin{array}{ccc}
                 -(-\partial_1^2+\omega^2) & 2 \ka \partial_1 & 0 \\
                 -2 \ka \partial_1 & -\partial_1^2+\omega^2 & -2 \omega \ka \\
                 0 & 2 \omega \ka & -\partial_1^2+\omega^2 \\
               \end{array}
             \right) \ .  \label{awl}
\end{eqnarray}
The $1$-loop correction to the effective action is then
\begin{eqnarray}
\Gamma_1= && \frac{\T}{4 \pi}\int_{-\infty}^{\infty} d \omega \bigg[-8 \ln \frac{\det[-\partial_1^2
+\omega^2+\rho'^2]}{\det[-\partial_1^2+\omega^2+ \ka^2 ]}+2 \ln
\frac{\det[-\partial_1^2+\omega^2+ 2 \rho'^2]}{\det[-\partial_1^2+ \omega^2+ 2 \ka^2]}\nonumber\\
&& - \ \ln \frac{\det^8 [-\partial_1^2+\omega^2+ \ka^2]}{\det^2 [
-\partial_1^2+\omega^2+ 2 \ka^2 ]\ \det^6 [-\partial_1^2+\omega^2]}+ \ln \frac{\det Q_{\omega}}
{\det Q^{(0)}_{\omega}}- \ln \frac{\det P_{\omega}}{\det Q^{(0)}_{\omega}} \bigg]\label{part}\ ,
\end{eqnarray}
where
\begin{eqnarray}
P_{\omega}=\left(
             \begin{array}{ccc}
               -(-\partial_1^2 + \omega^2)  & 0 & 0 \\
               0 & -\partial_1^2+\omega^2 & 0 \\
               0 & 0 & -\partial_1^2+\omega^2 \\
             \end{array}
           \right) \ . \la{puy}
\end{eqnarray}
Also,  $Q_{\omega}=Q^{(0)}_{\omega}+ \eta Q^{(1)}_{\omega}+ ...$, where $Q^{(1)}_{\omega}$ is the next
 to leading order coupled operator from (\ref{nxt})
\begin{eqnarray}
Q^{(1)}_{\omega}=\left(
               \begin{array}{ccc}
                   0 & Q_{12} & Q_{13} \\
                  Q_{21} & -\ka^2 \cosh(2 \ka \sigma) & Q_{23}\\
                   -Q_{13} & -Q_{23} & -\ka^2 \cosh (2 \ka \sigma)\\
                 \end{array}
             \right) \ .  \label{awl1}
\end{eqnarray}
Here
\bea
&&  Q_{12}=-\frac{\ka^2}{2}\sinh ( 2\ka \sigma) + i n [\frac{1}{\pi}-\frac{\ka}{2}\cosh (2 \ka \sigma)]\ , \\
&& Q_{21}=-\frac{\ka^2}{2}\sinh ( 2\ka \sigma) - i n [\frac{1}{\pi}-\frac{\ka}{2}\cosh (2 \ka \sigma)]
\ , \\
&& Q_{13}=-\frac{\ka \omega}{2}\sinh (2 \ka \sigma), \quad Q_{23}=-\omega [\frac{1}{\pi}+\frac{\ka}{2}\cosh (2 \ka \sigma)] \,
\eea
and we  performed the Fourier transform in $\s$, i.e.
replaced   $\partial_1 \rightarrow i n$, \ $n=0, \pm 1, ...$,
 as  appropriate for fluctuation fields which are  $2 \pi$ periodic in $\s$.

Our aim will be  to determine  the  1-loop correction to string energy to order $\eta$
by computing 
\be \la{eqf}
 \G_1 = \G_1^{(0)} + \G_1^{(1)} + \O(\eta^2) \ , \ \ \ \ \ \ \ \ \ \ \G_1^{(1)}= \O(\eta) \ . \ee
 As in \cite{tt} the first, second and
fourth terms in (\ref{part}) can be computed  to order $\O(\eta)$
using that
\begin{equation} \la{ll}
\ln \frac{\det [O^{(0)} + \eta  O^{(1)} ]}{\det\ O^{(0)}}=\eta\ {\rm Tr}[ (O^{(0)})^{-1} O^{(1)}]+\O(\eta^2) \ .
\end{equation}
While  in \cite{tt}   a similar    contribution to the effective action
 happened to vanish since it
  was proportional to the sum of squares of fluctuation masses,\foot{The mass
  sum rule
 implies the 1-loop  UV finitness of the superstring; it
  was proven in general for any string solution in \cite{dgt}.}
  here this  leading term is no longer  zero as in the present case 
 the  expansion  is around a nontrivial string background with different propagators
  for different string fluctuations.

\

In \rf{nxt}  we used the expansion \rf{rra}
of the solution $\r(\s)$ 
in small $\eta$. As discussed  in Appendices B and C, this
 expansion breaks down at the turning points where subleading terms are of 
 the same order as the leading term. 
As in the  computation  of  the  leading order in \cite{ft1},
 here we shall assume that one can ignore the 
 contributions from the turning points. 
 The  classical folded  string solution is built out of four 
 parts making up the closed string 
  (e.g., the expansion \rf{rra} used in \rf{nxt} is defined for 
 $0\leq \sigma \leq { \pi\ov 2} $).\foot{Note that for  the second and the fourth  $\sigma$ intervals
 where $\rho$ decreases we need to use the minus sign in (\ref{diff}).}
 The closed string fluctuations by definition must be 
   periodic in  $0\leq \sigma \leq 2 \pi$. 
   
 We shall assume that we can treat the problem ``piece-wise'' 
 also at the fluctuation level.  Direct implementation  of this 
  may effectively bring back the turning-point contributions.
  We shall  parametrize our current  lack of control of such terms by including 
  the  possible contribution with an arbitrary coefficient in the final result.
  
We shall split the 
integral over  $\sigma$ as follows
\begin{equation}
\int_0^{2 \pi} \frac{d \sigma}{2 \pi}\ \ \rightarrow\ \  \frac{1}{2 
\pi} \bigg[\int_0^{\frac{\pi}{2}} d \sigma+\int_{\frac{\pi}{2}}^{\pi} d 
\sigma+\int_{\pi}^{\frac{3 \pi}{2}} d \sigma+\int_{\frac{3 \pi}{2}}^{2\pi} d \sigma\bigg]\ . 
\end{equation} 
Considering the first interval $(0, {\pi \ov 2})$,  
the order $\eta$ contribution of the decoupled boson $\beta_u$ 
 in \rf{lag}  and \rf{lagr},\rf{nxt} can be obtained as
\begin{eqnarray}
 \bigg( \ln
\frac{\det[-\partial_1^2+\omega^2+ 2 \rho'^2]}{\det[-\partial_1^2+ \omega^2+ 2 \ka^2]}\bigg)^{(1)}
&=&- \frac{ \eta \ka}{\pi}\sum^\infty_{n=-\infty} \frac{1}{n^2+ \omega^2+ 2 \ka^2}\int_0 ^{\frac{\pi}{2}} \frac{d \sigma}{2 \pi}[ \pi \ka \cosh (2 \ka \sigma) -2]\nonumber\\
&=& -\frac{ \eta \ka}{4 \pi}\sum^\infty_{n=-\infty} \frac{\sinh 
(\pi \ka )-2}{n^2+ \omega^2+ 2 \ka^2}\ .  \la{tet}
\end{eqnarray}
Similarly,  for the fermionic contribution  (the first term in \rf{part}) 
we get
\begin{eqnarray}
\bigg( \ln
\frac{\det[-\partial_1^2+\omega^2+  \rho'^2]}{\det[-\partial_1^2+ \omega^2+  \ka^2]}
\bigg)^{(1)} 
&=&- \frac{ \eta \ka}{2 \pi}\sum^\infty_{n=-\infty} \frac{1}{n^2+ \omega^2+  \ka^2}\int_0 ^{\frac{\pi}{2}} \frac{d \sigma}{2 \pi}[ \pi \ka \cosh (2 \ka \sigma) -2]\nonumber\\
&=& -\frac{\eta \ka}{8 \pi}\sum^\infty_{n=-\infty} \frac{  \sinh ( \pi \ka )-2            }{n^2+ \omega^2+  \ka^2}
\ . 
\end{eqnarray}
For the coupled part one  finds 
\begin{eqnarray}
\bigg( \ln \frac{\det Q_{\omega}}
{\det Q^{(0)}_{\omega}}\bigg)^{(1)} 
&=&  \eta  \int_0^{\frac{\pi}{2}} \frac{d \sigma}{2 \pi} \Tr [(Q^{(0)}_{\omega})^{-1} Q_{\omega}^{(1)}]\nonumber\\
&=& \frac{  \eta\ka }{ \pi} \sum^\infty_{n=-\infty} \frac{(n^2+\omega^2)^2 -n^2 (n^2+\omega^2
 +\ka^2) \sinh ( \pi \ka)}{(n^2+\omega^2)^2 (n^2+\omega^2+ 4 \ka^2)}\ . \la{met}
\end{eqnarray}
The contributions of the  other three  intervals of $\sigma$ are  the same.


Collecting the above results 
 we observe that the final expression for the order $\eta$ term in the  effective action
 $\G^{(1)}_1$ 
  is UV finite. Moreover, the part that does not contain
   the $\sinh (\pi \ka)$ factor is IR finite, i.e. the non-trivial 
   potentially IR divergent 
   contributions of the  two unphysical $AdS_5$ massless modes $(\chi, \xi)$ 
   (time-like and longitudinal) 
    that appear in the coupled part of the fluctuation Lagrangian  
    cancel.\foot{Their flat-space contribution is cancelled against
     the conformal gauge ghost contribution.}

    Explicitly, integrating first over $\omega$ we obtain\footnote{Let 
    us mention that if we perform the sum over $n$ first, then using the 
    residue theorem in the integral over $\omega$ we arrive back at  the same  sum
    as below.} the order $\eta$ contribution to the $1$-loop effective action \rf{eqf}
    coming from (\ref{ll})
    as 
\begin{equation}
 \Gamma_1^{(1)} = -  \frac{\T \eta}{4 \pi} \sum^\infty_{n=-\infty} \Big[A_n + {C_n} 
  \sinh ( \pi \ka)\Big]\ , \la{gag}
\end{equation}
where
\begin{equation}
A_n= \frac{8 \ka}{\sqrt{n^2 +\ka^2}}-\frac{4 \ka}{\sqrt{n^2 +2 \ka^2}}-\frac{4 \ka}{\sqrt{n^2 + 4\ka^2}} \ ,  \label{owr}
\end{equation}
\begin{equation}
C_n= \frac{\ka}{2 n}+\frac{3 n}{4 \ka}- \frac{4 \ka}{\sqrt{n^2 +\ka^2}}+ \frac{2 \ka}{\sqrt{n^2 + 2 \ka^2}}- \frac{3 n^2 }{4 \ka \sqrt{n^2 + 4 \ka^2}}\ .
\end{equation}
The coefficient of the 
 part proportional to $\sinh (\pi \ka)$ given by  $\sum_n C_n$  is UV finite
 but formally has an IR singular contribution.\footnote{The IR singular contribution 
 goes away if one separates 
 it before doing the integral over $\omega$. Then 
 we get for the  large $\ka$ behaviour of  $\sum_n C_n$: \ \ 
$2\sum_{n=1}^{\infty}C_n =  \ka ({\ln \ka}-6 \ln 2 -\frac{3}{2}+
\gamma_{E}  )+O((\ka)^0)  . $
} 
 This term should be an artifact of our computational  procedure
  related to the problem 
 with  expansion in $\eta$ in \rf{rra} near the turning points (see Appendices B and C).
 Insisting on omitting the turning point contributions means that 
 we should drop this IR singular $\sim \sinh (\pi \ka)$ term, 
 and this is what we will do below.\foot{To justify the expansion in \rf{rra} we need to omit the turning point 
contribution. That can be done  by shifting the upper limit 
of the integration over $\s$ in \rf{tet}-\rf{met} to ${\pi \ov 2} - \epsilon$, \ 
$\epsilon \to +0$. Then the coefficient of $C_n$ will become 
$\sinh [(\pi- 2\epsilon) \ka]\sim \eta^{-1 + {2\ov  \pi} \epsilon}$
and thus is subleading compared to the contribution of order ${\cal O}(\eta^0)$.} 
 We believe that in a more systematic treatment that consistently 
 treats the turning point contributions such terms will be automatically absent
 (equivalently, in our present form of the  expansion,  such terms should  resum away,
  see also the discussion in  Appendix C).

Computing   the remaining  $\sum_n A_n$ contribution in \rf{gag},\rf{owr} using 
 the Euler-MacLaurin formula 
\begin{equation}
\sum_{n=1}^{\infty}f(n)= \int_1^{\infty}d n\ f(n) + \frac{f(1)+f(\infty)}{2}+\sum_{k=1}^{\infty}
 \frac{B_{2 k}}{(2 k)!}[f^{(2k-1)}(\infty)- f^{(2k-1)}(1)]\ ,
\end{equation}
we can extract its  large $\ka$ behaviour 
\begin{equation}
\sum_{n=-\infty}^{\infty} A_n = 12 \ka \ln 2 + \O(e^{-2 \pi \ka}) 
\ . 
\end{equation}
The contribution of this term in  $\Gamma_1^{(1)}$ in \rf{gag}  to the  energy \rf{loi}  is
then  (using \rf{spin})
\begin{equation}
E_1^{(1)}= - \frac{3 \ln 2 }{\pi }{\ka \ov \k}  \ \eta 
\ . \la{hja}
\end{equation}
Let us now include  the  $O(\eta^0)$ contribution
to $\G_1$ \rf{eqf} coming  from the third and fifth
 terms in (\ref{part}). 
 Since $Q^{(0)}_{\omega}$ in \rf{awl} has no $\sigma$ dependence, its
 functional determinant
 \begin{equation}
\det Q_{\omega}^{(0)}= - \det^2 (-\partial_1^2 +\omega^2)\ \det (-\partial_1^2 +\omega^2 +
 4 \ka^2)\ 
\end{equation}
  can be easily computed as a product over integer $n$ of a matrix
 determinant (after 
  $\partial_1  \rightarrow i n$). 
Since $\det P_{\omega}= - \det^3 (-\partial_1^2 +\omega^2)$ we
may  write the relevant contribution
  from (\ref{part}) as
\begin{equation}
\Gamma^{(0)} _1= - \frac{\T}{4 \pi}\int_{-\infty}^{\infty} d \omega \ln \frac{\det^8 (-\partial_1^2+\omega^2+
 \ka^2)}{\det^2 (-\partial_1^2 +\omega^2 +  2 \ka^2)\ \det^5 (-\partial_1^2 +\omega^2)
 \ \det (-\partial_1^2+\omega^2 +   4 \ka^2)}\ . \la{ghj}
\end{equation}
Since  $\ln \det (-\partial_1^2 + \omega^2 + \kappa^2)= \sum^\infty_{n=-\infty} \ln(n^2+\omega^2 +\kappa^2)$,
 doing the integral over $\omega$ we finally obtain the $1$-loop correction to
 the string  energy to order
 $\O(\eta)$ as
\begin{equation}
E_1^{(0)} = \frac{1}{2 \kappa} \sum_{n=-\infty}^{\infty} \bigg[\ 2 \sqrt {n^2+2 \ka^2}+
 \sqrt{n^2+ 4 \ka^2}+5 \sqrt{n^2}-8 \sqrt{n^2+ \ka^2}\  \bigg]\ .\la{suma}
\end{equation}
This turns out to be  a direct generalization of the
leading-order result of \ci{ft1}  where $\k$ should be replaced by $\ka$
in the fluctuation  mass terms (but not in the overall $1 \ov \kappa$  factor due to $t =\kappa \tau$). 

Using again the Euler-MacLaurin formula to transform the sum into an integral
we find\foot{Note that the sum of \rf{owr} is  minus the derivative over $\ka$ of the sum 
in \rf{suma}, which explains why the corresponding coefficients  are closely related.}
\begin{equation}
E_1^{(0)}= \frac{1}{\kappa}\ \Big[- 3 \ln 2 \  \ka^2 \  - \frac{5}{12} \
 + \O(e^{-2 \pi \ka})\Big]\ . \la{ty}
\end{equation}
This   is, of course,  in agreement with the result of \ci{ft1} and
also, for the  subleading term,
 with ref. \cite{sak}.\foot{Ref.\ci{sak}
 considered, following \ci{ft1},
  the formal sum \rf{ghj} with $\ka \to \k$   and with the $n=0$ term omitted
 (this term  was omitted in \ci{ft1} since there it was subleading
  in the infinite  $\k$ limit).
 As a result, the expression in \ci{sak} contained an extra (minus ``zero mode'')
term  $3-\sqrt 2 $. 
Note that the coefficient in the  exponent of the leading  $ e^{-2 \pi \ka}$ exponential 
correction  \cite{sak}
is determined by the mass of the lightest  mode -- in the present  case of the fermionic mode.
The exponential term in the square bracket 
has also a prefactor of  $\sqrt{\kappa_0}$. 
}

As we shall argue in Appendix C, an additional  contribution
that may  come from near turning point regions 
can be  parametrized as follows:
\be \la{tuu}
\Gamma_1^{(2)}=  {\ck \ov \pi}  \ \k_0\ \T \ ,  \ \ \ \ \ \ \ \ \ 
 E_1^{(2)}  = {\ck \ov \pi} \ {\k_0 \ov  \k }  \ , 
\ee 
where $\ck$ is an undetermined constant (we included factor of $\pi$ for convenience).

Inverting the relation between  $\S$ and $\eta$ in  (\ref{spin}) to order
$\O(\eta)$ we get
\begin{equation}
\eta= \frac{2}{\pi \mathcal{S}}- \frac{ \ln ( 8 \pi \mathcal{S})  -3 }{\pi^2 \mathcal{S}^2}+...\ ,
\end{equation}
which,  plugged into (\ref{appa}),  gives\footnote{Note that  in the expression for the  energy in \rf{loi}  we need to
    keep $\kappa$ to order $\O(\eta)$ to get the  correction in $E_1$ also
    to order $\eta \sim {1 \ov   \S}$.}
\begin{equation}
\kappa=\frac{\ln (8 \pi \mathcal{S})}{\pi}   -\frac{1}{2 \pi^2 \mathcal{S}}   + ... 
\ , \ \ \ \ \ \ \ \ \ \ 
\kappa_0=\frac{\ln (8 \pi \mathcal{S})}{\pi} +  \frac{ \ln ( 8 \pi \mathcal{S})  -3 }{2\pi^2 \mathcal{S}} +   ... \ , 
\ \ \ \ \ \   {\kappa_0 \ov \kappa} = 1   + {1 \ov 2 \pi \S} + ...  \ . 
\end{equation}
This means that the  dominant term in \rf{ty} is the $ - 3 \ln 2\ { \ka^2 \ov \k}$ one:
the other terms
$ -\frac{5}{12 \k } \sim  {1\ov  \ln \S }$ and $ {\sqrt{\ka}  \ov \k} e^{-2\pi \ka} \sim
 { 1 \ov \sqrt{ \ln \S}\  \S^{ 2}}$  should  be subleading and should be ignored in the approximation we
 considered above  where we  dropped terms of higher order in $\eta$ at  earlier stages; these terms are expected to  cancel out in a more systematic treatment.\foot{The
 role of these  subleading terms  and their possible resummation  remains to be studied
 in more detail.}

Thus we  find
  that the $1$-loop correction to the folded string energy to order
$\O(\frac{\ln^2 \mathcal{S}}{\mathcal{S}^2})$ can be  written in
the same form as the classical energy
\rf{EmS1}, i.e. as
in \rf{exx},\rf{xax}
\begin{equation}\la{oka}
E_1= b_0 \ln \mathcal{S}+ b_c + \frac{b_{11} \ln \mathcal{S}+b_{10}}{\mathcal{S}}+ \O(\frac{\ln^2
 \mathcal{S}}{\mathcal{S}^2}) \ .
\end{equation}
The contribution from $E_1^{(0)}$ in \rf{ty} to the $1$-loop coefficients is
\begin{equation}\la{uuu}
b_0^{(0)}=- \frac{3 \ln 2 }{\pi}, \ \   \quad b_c^{(0)}= - \frac{3 \ln 2 }{\pi} \ln 8 \pi,
\ \  \quad b_{11}^{(0)}= - \frac{3 \ln 2}{\pi^2},\ \  \quad b_{10}^{(0)}=-\frac{3 \ln 2}{ \pi^2}
\big( \ln 8 \pi - \frac{5}{2} \big) \ ,
\end{equation}
while  $E^{(1)}_1$ in \rf{hja}  and  $E_1^{(2)}$  in \rf{tuu}   contribute  as
\begin{equation}
b_{10}^{(1)}= - \frac{6 \ln 2}{\pi^2} \ , \ \ \ \ \ \ \ \ \ \ \ 
b_{c}^{(2)}=      {1 \ov \pi}\ \ck  \ ,   \ \ \ \ \ \ 
b_{10}^{(2)}=   {1 \ov 2\pi^2 }\ \ck   \  .  
\end{equation}
Thus  finally we obtain for the full $1$-loop coefficients  
\bea\la{uuu2}
&&b_0=- \frac{3 \ln 2 }{\pi}\ , \ \ \ \ \ \ \   \quad b_c={1 \ov \pi}   \big( 
 - {3 \ln 2 } \  \ln 8 \pi + {\ck} \big) \ , \\
&&  
 b_{11}= - \frac{3 \ln 2}{\pi^2},\ \  \ \ \ \ \  \quad b_{10}={1 \ov 2\pi^2}  \Big[ -{6 \ln 2}\ 
\big( \ln 8 \pi - \frac{1}{2} \big)  + {\ck  } \Big] \ .\la{uuu3}
\eea
The functional and reciprocity relations in \rf{fun},\rf{VV}
  at strong coupling are (see discussion in Appendix E)
\be
&&  \bar f_{11} = \ha \bar f^2 \ , \ \ \  \ \ \ \ \bar f_{10} = \ha   \bar f \bar f_c \ , \ \ \ \ \ \ \ \ \ \ \
\bar f \equiv   { f \ov \sql}\ , \ \   \bar f_{c}\equiv  { f_{c} \ov \sql}\ ,  \ \
\bar f_{1k}\equiv  { f_{1k} \ov \l} \ ,
\la{gho}\\
&&
\bar f  =  a_0 + {b_0 \ov \sql } + ...\ , \ \ \ \ \ \
 \bar f_{c}=  a_{c} + {b_{c} \ov \sql } + ... \ ,
\ \ \ \ \ \
\bar f_{1k}=  a_{1k} + {b_{1k} \ov \sql } + ...\ .
\la{joq} \ee
They imply that the coefficients in \rf{oka}   should  obey 
(see \rf{amm},(\ref{aml}))
\begin{equation}
b_{11}= a_0 b_0\ ,\ \ \ \ \    \qquad b_{10}=\ha ({a_0 b_c+ b_0 a_c}) \ . \la{www}
\end{equation}
Recalling the  values of the leading coefficients at the classical level in  \rf{EmS1}
\begin{equation}
a_0=\frac{1}{\pi}\ ,\ \ \ \  \qquad a_c= \frac{1}{\pi} ( \ln 8 \pi-1  )  \ ,\la{wer}
\end{equation}
we  see that the relations  \rf{www}  are indeed satisfied by the expressions in \rf{uuu2},
i.e. the functional  and the  reciprocity relations
appear to  apply   also including string 1-loop corrections. Note that 
this is true for any value of teh undetermined  coefficient $\ck$.

Needless to say, it would be interesting to generalize the 1-loop  computation
of this section  and the  check of reciprocity
to the case of   non-zero $J$  and to attempt to relate  the  strong-coupling version
of the reciprocity discussed in Appendix E   to its weak-coupling finite
twist one in \rf{MVV}.

\section*{Acknowledgments }

We are grateful to  F. Alday, G. Arutyunov, B. Basso,  L. Dixon, S. Frolov, N. Gromov,
 G. Korchemsky, M. Kruczenski, T. McLoughlin,  A. Rej,  R. Roiban,  D. Seminara  and S. Theisen
 for useful discussions. 
 We thank G. Korchemsky  and  R. Roiban for very  helpful comments on the draft. 
 A.A.T. thanks  I.Y. Park  for
 a collaboration on this  topic back in 2005.
The research of V. Forini is supported by the SFB 647
 `Space-Time-Matter' grant and by the Alexander von Humboldt foundation.
   A.T. was supported in part by NSF under grant PHY-0653357.
Part of this work was done while V.F. and A.A.T. were participants
of the GGI Florence workshop ``Non-Perturbative
Methods in Strongly Coupled Gauge Theories" and we
 thank the Galileo Galilei Institute for Theoretical Physics for the hospitality
 and the INFN for partial support.

\bigskip
\bigskip

\section*{Note Added }
As we have learned (N. Gromov, private communication)
an independent way of evaluating the 1-loop correction  to the folded string energy 
based on the algebraic curve approach to extracting fluctuation frequencies \ci{gvs}
  leads to the value
$\ck= 6 \ln 2 + \pi  $ for 
the undetermined coefficient in \rf{kk},\rf{l},\rf{lll},\rf{tuu},\rf{uuu2},\rf{uuu3}.
This  $\ck$ contribution  changes the 1-loop coefficient in \rf{kk} from $3 \ln 2 $ to 
$ - 3 \ln 2 -\pi$. 
It is interesting to note that  for $\ck= 6 \ln 2 + d $, where $d$ is a
constant not involving $\ln 2$, the coupling
redefinition
$\sql \to  \sql  + 3 \ln 2 $ (suggested  to us by G. Korchemsky)
removes  all  $\ln 2$ terms  from the leading  1-loop coefficients in
\rf{faj}--\rf{lll},  just as it did in the  cusp anomaly
coefficient in \ci{bkk}. Namely, then we get 

$
f= {\sql \ov \pi} + O( {1 \ov \sql}),\ \ \
f_c=  {\sql \ov \pi} [\ln  {8 \pi\ov \sql} -1]+ {d \ov \pi  } + O( {1 \ov \sql}),\
\ \
f_{10}= {\lambda \ov 2 \pi^2} [\ln  {8 \pi\ov \sql} -1] + {d \sql \ov 2 \pi^2}
 + O(\lambda^0).$

\bigskip

\appendix
\subsection*{Appendix A:  Comments on conformal algebra realizations }
\refstepcounter{section}
\def\theequation{A.\arabic{equation}}
\setcounter{equation}{0}

Starting with a conformal theory in $R^{1,3}$ with the standard $SO(2,4)$
conformal group generators $P_m, M_{mn}, K_m, D$ ($m,n=0,1,2,3$) one may define the collinear
$SL(2, R)$  subgroup as generated by  the following
light-cone components  \ci{bra}:
\be
&&L_+ \equiv  -i P_+,\ \ \ \ \ L_- \equiv  {i \ov 2} K_-,\ \  \ \  L_0 \equiv
{i \ov 2} ( D + M_{-+}), \\
&& [L_0, L_\pm]= \pm L_\pm, \ \ \ \ \  \   \ [L_+, L_-] = - 2 L_0 \ .
\ee
If the eigenvalue of $D$  is dimension $\Delta$  and the eigenvalue of
   $M_{-+}$  -- the collinear projection of the  Lorentz spin  $S$, then the eigenvalue of
$L_0$ is the conformal spin $s= \ha ( \Delta + S)$. The corresponding
quadratic Casimir operator is $C^2= s(s-1)$.

At the same time, in the   $R^{2,4}$
embedding representation of the global $AdS_5$ space   the generators  $\Si_{MN}$
($M,N=0,1,2,3,4,5$  with   the signature $-++++-$)
of  $SO(2,4)$   linearly realised on the embedding coordinates $Y^M$
 can be related  to
 the standard boundary  conformal group generators  as (see, e.g.,  \ci{my})
  \be \la{ni}
\Si_{mn} = M_{mn} \ , \ \ \ \ \ \
 \ \Si_{m 4} = \ha (K_m - P_m)\ , \ \ \ \ \ \
 \    \Si_{m 5} = \ha (K_m + P_m) \  , \   \ \ \ \
  \ \ \Si_{54} = D\  .  \ee
Then  the standard spin is    $S = \Si_{12} = M_{12}$,
the conformal  spin  is   $S' = \Si_{34} = \ha (K_3- P_3)$,
 and  the conformal   energy  is
   the  rotation generator  in the $05$ plane, i.e.
   the global $ AdS_5$ energy, \  $E= \Si_{05} = \ha (K_0 + P_0)$.

 In general, the  energy $E$ of a string state   in global $AdS_5$ space
  with boundary $R \times S^3$  should be equal to the energy of
  the corresponding SYM  state
 on $R\times S^3$.
 Through radial quantization (and analytic continuation)
  this  state may be
 associated to a local operator in $R^{1,3}$  that creates it.
 The $AdS_5$ energy $E=\Si_{05}$ or conformal Hamiltonian
  generates an $SO(2)$ subgroup  while the dilatation
  operator $D=\Si_{54}$
  generates an $SO(1,1)$ subgroup  of $SO(2,4)$.\foot{Their
    eigenvalues happen to be  the same
  since the two  representations (the unitary one
  classified by $SO(4) \times SO(2)$  and the one
  classified by  $SO(4) \times SO(1,1)$)
   are related by a global $SO(2,4)$
  similarity transformation (see, e.g., \ci{dol}).}
  After the Euclidean continuation of the embedding coordinate
 $Y_0\to  i Y_{0E}$  (to allow for the  mapping from  $R \times S^3$ to $R^4$)
 one may exchange   $Y_{0E}$ with $Y_4$ which
exchanges the generator  $\Si_{54} = D$ with $E= \Si_{05} = \ha ( P_0 + K_0)$.

 To  relate the $SO(1,2)$ subgroup of $SO(2,2)$  which is a symmetry of
global  $AdS_3$  subspace of $AdS_5$   where the
folded spinning string  is moving
to  the collinear $SL(2, R)$ subgroup  classifying the operators like
$\tr ( \Phi \DD^S_+ \Phi)$
 one is  also to perform
  an additional analytic continuation
   that  interchanges the euclidean (12) plane
with the hyberbolic $(+-)$ plane.
Since different choices are formally related via $SO(2,4)$
transformations and a re-identification of the generators one may
expect that the two representations should be equivalent.\foot{The formal relation
 can be achieved  by a  continuation to euclid:
by replacing null direction  like $x_0+ x_3$  with  a complex one $x_1 + i x_2$,
i.e. replacing the operator $\tr ( \Phi \DD_+^S \Phi)$  with
$\tr ( \Phi \DD_*^S \Phi)$, where $\DD_* = \DD_1 + i \DD_2 $.}

Still, the representations of $SO(2,2)$  or string states in $AdS_3$
are  naturally labeled by $(E,S)$,
 and the relation to $SO(1,2)$  labels $\ha (E+S)$ does not appear to be
 natural, {\it unless} one is interested in the  large  spin expansion 
  (see in this connection
  \ci{krt,iktt}).
  That relation  may possibly be made more explicit  by choosing a different set of coordinates
 in global $AdS_5$ in which the boundary is not $R \times S^3$ but $AdS_3 \times S^1$
  (see \ci{am2}  where such  coordinates in the boundary theory where used
 to explain the leading $E \sim \ln S$ behaviour).

 Let us mention also that  the relation \rf{faw} or  $E-S = {\rm f} (E+S)$ is
 reminiscent of a light-cone gauge expression, where ${\rm f}$ would be a
 light-cone Hamiltonian (cf. \ci{mtth,afr,krt}).



\appendix
\refstepcounter{section}
\def\theequation{B.\arabic{equation}}
\setcounter{equation}{0}

\subsection*{Appendix B:  Review of folded string solution with $J=0$}

 In this Appendix we review the folded spinning string
 solution in $AdS_3$  \ci{dev,gkp}  and consider its large spin expansion
 (see also \cite{ft1}).

 The solution is given by
\begin{equation}\la{so}
 t= \kappa \tau, \quad \phi= w \tau, \quad \rho=\rho(\sigma) \ , \ \ \ \ \ \ \
ds^2= -\cosh^2 \rho\ dt^2 + d \rho^2 + \sinh^2 \rho\ d \phi^2
\ , \end{equation}
where  $\rho(\sigma)$ satisfies
\begin{equation}
\rho' = \pm \kappa \sqrt{1- \eta \sinh^2 \rho}  \label{diff} \ .
\end{equation}
Here $\rho$ varies from $0$ to its maximal  value $\rho_0$ related to the parameter $\eta$ by
\begin{equation}
\coth^2 \rho_0 = \frac{w^2}{\kappa^2}\equiv 1+ \eta \ .
\end{equation}
The solution in the interval $0 \leq \sigma \leq \frac{\pi}{2}$ with the
initial condition $\rho(0)=0$ is
\begin{equation}
\sinh \rho = \frac{1}{\sqrt{\eta}}\ {\rm sn} \big[\kappa \sqrt{\eta}\ \sigma, -\frac{1}{\eta}\big]\ , 
 \ \ \ \ \ \  0 \leq \sigma \leq \frac{\pi}{2}\ . 
  \label{sol}
\end{equation}
The condition satisfied at the turning point
 $\rho_0$ at $\sigma=\frac{\pi}{2}$ is  $\rho'(\frac{\pi}{2})=0$.
  To construct the full ($2 \pi$ periodic)
  folded closed string solution
  one should  glue together four such functions  on $\pi \ov 2$ intervals
  to cover the full $0\leq \sigma \leq 2 \pi$ interval;
  e.g., for ${\pi \ov 2} < \s < \pi $ we have
  \begin{equation}
\sinh \rho = \frac{1}{\sqrt{\eta}}\ {\rm sn} \big[\kappa \sqrt{\eta}\ ( \pi -\sigma), -\frac{1}{\eta}\big]\ , \ \ \ \ \ \ \ \ {\pi \ov 2}  \leq \sigma \leq \pi \ .
  \label{sols}
\end{equation}
  The expressions for the parameter
    $\kappa$,  the energy and the spin in terms of $\eta$ are
\begin{equation}
\kappa= \frac{1}{\sqrt{\eta}}\ _2 F_1(\frac{1}{2},\frac{1}{2};1;-\frac{1}{\eta}),
\quad \mathcal{E}=\frac{1}{\sqrt{\eta}}\ _2 F_1(-\frac{1}{2},\frac{1}{2};1;- \frac{1}{\eta}),
\quad \mathcal{S}=\frac{\sqrt{1+\eta}}{2 \eta \sqrt{\eta}}\ _2 F_1(\frac{1}{2},\frac{3}{2};2;- \frac{1}{\eta}) \la{defa}
\end{equation}
In this paper we are interested in the large spin or long string limit,
 i.e. small $\eta$ expansion.
 Since in section 3  we compute $1$-loop correction only
 to order
 $\frac{1}{\mathcal{S}}$ we will
 only need expansions to order $\O(\eta)$.
 Expanding $\kappa,\ \E,\ \S$ in small $\eta$ we obtain~\footnote{
 Let us note that these expansions were  found  using pre-Mathematica 6
 versions  of Mathematica  (Mathematica 6  apparently has some bug 
 leading to inconsistent expansions 
 for some  elliptic and hypergeometric functions).}
\begin{equation}
\kappa= \ka - \frac{\eta}{4 \pi}(\pi \ka - 2)+\O(\eta^2) \ , \ \ \ \ \ \ \ \ \ \ \
\ka\equiv       { 1 \ov \pi} \ln { 16 \ov \eta} \ ,  \label{kappa}
\end{equation}
\begin{equation}\la{eqt}
\mathcal{E}=\frac{2}{\pi \eta}+\frac{\pi \ka +1  }{2 \pi}- \frac{\eta}{32 \pi}( 2 \pi \ka -3 )+\O(\eta^2)
\ , \end{equation}
\begin{equation}
\mathcal{S}=\frac{2}{\pi \eta}- \frac{\pi \ka -3}{2 \pi}-\frac{\eta}{32 \pi}(
2 \pi \ka + 13  )+\O(\eta^2)\ .   \label{spin}
\end{equation}
 Expanding the solution (\ref{sol}) in small $\eta$ we obtain, for $0 < \sigma < {\pi \ov 2}$,
\begin{equation}
\sinh \rho= \sinh (\kappa \sigma) - \frac{\eta}{8} \big[ \sinh (2 \kappa \sigma)
- 2 \kappa \sigma \big]
\cosh (\kappa \sigma) +\O(\eta^2)\ , \la{rraa}
\end{equation}
or,  using \rf{kappa},
\begin{equation}
\sinh \rho =  \sinh (\ka \s)-  \frac{\eta}{8 }\big[ \sinh (2 \ka \s)-
{\textstyle {4\ov \pi} } \sigma \big] \cosh(\ka \s)+\O(\eta^2) \ .\la{rra}
\end{equation}
To leading order when $\eta\to 0, \ \ka \to \infty$ 
the string  touches the boundary of $AdS_5$ ($\rho_0=\infty$)
and  the solution can be approximated (away from the turning
points) by simply 
 $\rho=\ka \sigma$.
 This limiting case  proved to be a useful framework
  for computing 1- and $2$-loop string corrections \cite{ft1,ftt,rtt,rt}.
At the  next order in small $\eta$ expansion
the ``ends'' (turning points)  of the string are close to the boundary but no longer touch it.

We  should add a word of caution about the use  of the formal expansion in
\rf{rraa} or \rf{rra}. Notice that it goes in powers of $\eta$ with coefficients
containing $\ln \eta$  and  is not, strictly speaking, valid close enough
to the turning points. Indeed, for $\s= {\pi \ov 2} $ we have
$\sinh (2\ka \s)= \sinh(\pi  \ka) \approx \ha e^{\pi \ka} \sim  \eta^{-1}$ and similarly
$\cosh (\ka \s)= \sinh ({ \pi \ov 2}\ka ) \sim  \eta^{-1/2}$. Hence at  the turning point  
 the order $\eta$ term in \rf{rra} goes actually as $\eta^{-1/2}$, 
 i.e. is of the same order as the leading term $\sinh (\ka \s)$.
If $\sigma$ is slightly away from  the turning point the
subleading terms are smaller than the leading term but then the expansion
and the contributions to the energy need to be resummed. 
The resummation at the level of the string profile $\rho(\sigma)$ is completely 
equivalent to its expansion 
near the turning point $\sigma={\pi\ov 2}$, as will be  discussed in Appendix C.

In section 3  we ignored 
the regions near the turning points  
 and 
thus uses the formal expansion \rf{rra} to compute it.
Similar assumption was made in \ci{ft1} in the computation of the 1-loop shift of
the coefficient of the $\ln \S$ term in the energy (where it was indeed justified). 
A reason behind this assumption is that the masses of string fluctuations in
\rf{as}  depend on $\rho'^2$  which is small near the turning points where
$\rho'=0$.  Thus the correction to the leading result of \ci{ft1}
should mainly come from the ``internal'' parts of the $\s$-interval, where the 
expansion \rf{rra}  is justified.
We shall comment  more on this  point 
in Appendix C. 
In section 3 we included  a possible term 
which we may thus miss  with an aribitrary coefficient yet to be deterined. 
 

For the computation of the $1$-loop correction in section 3  we will
 need the following expansions
\be
& &\rho'^2 = \ka^2 - \frac{\eta}{2 \pi}\ka [\pi \ka \cosh (2 \ka \s)- 2] +
 \O(\eta^2)\ ,\la{roo}
\\
\kappa \sinh \rho &=& \ka \sinh(\ka \s)
+  \frac{\eta}{8 \pi} \big\{ 4 \ka \sigma  \cosh(\ka \s)\nonumber\\
& & - [ \pi \ka \cosh(2  \ka \s) +3 \pi \ka -4 ]\sinh(\ka \s)\big\}+\O(\eta^2) \ , \\
w \cosh \rho &=& \ka \cosh(\ka \s)
+\frac{\eta}{8 \pi} \big\{4  \ka \sigma  \sinh(\ka \s)\nonumber\\
& &- [ \pi \ka \cosh (2 \ka \s) - 3 \pi \ka -4   ]\cosh(\ka \s)\big\}+\O(\eta^2)\ .
\ee
Notice again that these expansions are formally invalid at the turning point
(but are  justified away from it). This is evident, e.g., in
 \rf{roo} where the leading term does not vanish at the turning point where one should have $\rho'=0$.
 What happens is that at $\s= {\pi \ov 2}$  the leading term $\ka^2$ gets 
   cancelled agianst the sum of  subleading terms
 which all are of the same order, i.e. are proportional to $\ka^2$
 (see  Appendix C).

The masses appearing in the fluctuation Lagrangian in section 3
are then expanded as follows
\be
&&\mu_t^2= \ka^2 - \frac{\eta \ka}{ \pi} \big[  \pi \ka  \cosh(2 \ka \s)
- \ha \pi \ka -1 \big]+\O(\eta^2)\ , \\
&&\mu_{\phi}^2= \ka^2 - \frac{\eta \ka}{ \pi} \big[ \pi \ka \cosh(2 \ka \s)
+ \ha \pi \ka -1
\big]+\O(\eta^2)\ , \\
&&\mu_{\rho}^2=-\eta \ka^2 \cosh(2 \ka \s)+\O(\eta^2)\ . \la{maal}
\ee


\appendix
\subsection*{Appendix C:  
Resummation of ``long string'' expansion 
near the turning 
points}
\refstepcounter{section}
\def\theequation{C.\arabic{equation}}
\setcounter{equation}{0}

Continuing the discussion of the previous Appendix B here 
 we will 
 show that it is possible to resum systematically the terms $\sim e^{2\,n\,\kappa_0\,\sigma}$
appearing in the formal $\eta$ expansion of $\rho(\sigma)$ 
in \rf{rraa}.
 These terms are potentially dangerous since 
they scale at the turning point $\sigma=\pi/2$ as $e^{n\,\pi\,\kappa_0} = (16/\eta)^n$ and 
spoil the perturbative expansion.

In the first  ``quarter-string''  interval 
 $0\le\sigma\le\frac{\pi}{2}$,
 the function $\rho(\sigma)$ obeys the differential equation \rf{diff}   with plus sign and 
 $\rho(0)=0$.
Its formal expansion in powers of $\eta$ (treating $\kappa_0$ in \rf{kappa}
as a constant parameter)
 reads (cf. \rf{rraa}) 
\ba
&&\rho(\sigma) = \kappa_0\,\sigma+\Big[\frac{\sigma }{2 \pi }-\frac{1}{8} \sinh \left(2 
\kappa_0\,\sigma\right)\Big] \eta \no\\
&&+\ \Big[-\frac{\cosh \left(2 \kappa_0\,\sigma\right)
   \sigma }{8 \pi }-\frac{13 \sigma }{64 \pi }+\frac{1}{16} \sinh \left(2 \kappa_0\,\sigma\right)+
   \frac{1}{256} \sinh \left(4 \kappa_0\,\sigma\right)\Big]
   \eta ^2\nonumber \\
&& + \ \Big[-\frac{\sinh \left(2 \kappa_0\,\sigma\right) \sigma ^2}{16 \pi ^2}+\frac{29 \cosh \left(2 
\kappa_0\,\sigma\right) \sigma }{256 \pi
   }+\frac{\cosh \left(4 \kappa_0\,\sigma\right) \sigma }{128 \pi }+\frac{23 \sigma }{192 \pi 
   }   \la{exxq}\\
&& -\frac{1}{128} \cosh \left(2 \kappa_0\,\sigma\right) \sinh
   \left(2 \kappa_0\,\sigma\right)-\frac{\cosh \left(4 \kappa_0\,\sigma\right) \sinh \left(2 
   \kappa_0\,\sigma\right)}{3072}-\frac{125 \sinh \left(2 \sigma 
   \kappa _0\right)}{3072}\Big] \eta ^3+O\left(\eta ^4\right).\nonumber
\ea
Since here, in fact,  $\kappa_0 = -\frac{1}{\pi}\ln\frac{\eta}{16}$ the hyperbolic functions 
potentially reduce the true order in the $\eta$ expansion.
The dangerous terms can be easily identified by setting
\be
\tt = e^{\kappa_0\,\sigma},\qquad 
\quad \ff(\tt) \equiv  \rho\left(\frac{\ln \tt}{\kappa_0}\right),
\ee
and neglecting exponentially suppressed terms in the above expansion. The next-to-leading (NLO) 
result which is correct at $\oh{\eta}$ can be written
\ba
\ff_{_{\rm NLO}}(\tt) &=& \ln \tt
-\frac{\eta  \tt^2}{16}+\frac{\eta ^2 \tt^4}{512}-\frac{\eta ^3 \tt^6}{12288}+
\frac{\eta ^4 \tt^8}{262144}+\cdots \la{qqq} \\
&&  
+ \ \left[
\frac{\eta  \tt^2}{32}-\frac{\eta ^2 \tt^4}{512}+\frac{\eta ^3 \tt^6}{8192}+\cdots + 
\frac{\ln t}{2 \pi  \kappa _0}\left(
1-\frac{\eta  \tt^2}{8}+\frac{\eta ^2 \tt^4}{128}-\frac{\eta ^3 \tt^6}{2048}+\cdots
\right)
\right]\,\eta + \dots~.\nonumber
\ea
All terms $(\eta\,\tt^2)^k$ are $\oh{1}$ at $\sigma={\pi\ov 2}$ and the 
 above infinite series need resummation.
This can be accomplished as follows. 
Introducing  $h(\tt) \equiv  \ff(\tt)-\ln \tt$ we get 
\be
1+\tt\,h'(\tt) = \frac{\kappa}{\kappa_0}\sqrt{1-\eta\left(\frac{\tt\, e^h-\tt^{-1} e^{-h}}{2}\right)^2}
\ , \ \ \ \ \ \ h(\tt) = \ff(\tt)-\ln \tt\ . 
\ee
Taking the large $\tt$ limit we arrive at the following equation for the leading order term $h_{\rm LO}$ 
\be
1+\tt\,h_{\rm LO}'(\tt) \simeq  \sqrt{1-{\textstyle \frac{1}{4}} \eta\tt^2e^{2\,h_{\rm LO}(\tt)}}\ , 
\ee
which can be integrated and gives
\be
h_{\rm LO}(\tt) = -\ln\left(1+\frac{\eta\,\tt^2}{16}\right)\ .
\ee
As a check we can reexpand and find indeed
\be
h_{\rm LO}(\tt) =  -\frac{\eta  \tt^2}{16}+\frac{\eta ^2 \tt^4}{512}-\frac{\eta ^3 \tt^6}{12288}+
\frac{\eta ^4 \tt^8}{262144}-\frac{\eta ^5 \tt^{10}}{5242880}+{\cal O}\left(\tt^{11}\right)
\ee
which are the leading terms in $\ff(\tt)$. 

The NLO approximation $h_{_{{\rm NLO}}}(\tt)$ is simply obtained by including an extra piece in 
the square root and taking into account that
the ratio $\kappa/\kappa_0$ has a non trivial expansion in $\eta$, 
\be
1+\tt\,h'_{{_{_{\rm NLO}}}}(\tt) = \frac{\kappa}{\kappa_0} \sqrt{1-\eta\frac{\tt^2}{4}
 e^{2\,h_{_{_{\rm NLO}}}(\tt)}+\frac{\eta}{2}}\ .
\ee
Integrating  this equation, substituting  the necessary terms in the expansion of $\kappa/\kappa_0$, 
and neglecting  all NNLO terms  we then get  
\be\la{iii}
h_{_{_{\rm NLO}}}(\tt) = -\ln\left(1+\frac{\eta\,\tt^2}{16}\right) + \eta\,\frac{\ln t}{2\pi\,\kappa_0}+\frac{\eta^2\,\tt^2}{32}\,\frac{1}{1+\frac{\eta\,\tt^2}{16}}\,
\left(1-\frac{2\,\ln t}{\pi\,\kappa_0}\right).
\ee
The  expansion of \rf{iii}  in powers of $\eta$ gives
\ba
h_{_{_{\rm NLO}}}(\tt) &=& 
\left(\frac{\ln t}{2 \pi  \kappa _0}-\frac{\tt^2}{16}\right) \eta 
+\left(\frac{\tt^4}{512}-\frac{\tt^2\ln t}{16 \pi  \kappa _0}+\frac{\tt^2}{32}\right) \eta
   ^2\no \\
&& + \left(-\frac{\tt^6}{12288}+\frac{\tt^4 \ln t}{256 \pi  \kappa _0}
-\frac{\tt^4}{512}\right) \eta ^3 + \left(\frac{\tt^8}{262144}-\frac{\tt^6 \ln t}{4096 \pi  \kappa
   _0}+\frac{\tt^6}{8192}\right) \eta ^4\nonumber\\
&& + \left(-\frac{\tt^{10}}{5242880}+\frac{\tt^8 \ln t}{65536 \pi  \kappa _0}
-\frac{\tt^8}{131072}\right) \eta ^5+\cdots~,  
\ea
which  agrees indeed with the expansion of $\ff_{_{_{\rm NLO}}}$ in \rf{qqq}.

\medskip
In terms of the string profile
 $\rho(\sigma)=\ff( e^{\ka \s}) = h( e^{\ka \s}) - \ka \s  $ 
this result can be written as
\ba
\label{eq:resummation}
\rho_{_{_{\rm NLO}}}(\sigma) &=&  \kappa_0\,\sigma  -\ln\Big[1 + \left(\frac{\eta}{16}\right)^{1-2\sigma/\pi}\Big] 
 \nonumber \\
&& +\ \eta\,\Big[\frac{\sigma}{2\,\pi}+\frac{1}{2}\,\frac{\left(\frac{\eta}{16}\right)^{1-2\sigma/\pi}}{1+\left(\frac{\eta}{16}\right)^{1-2\sigma/\pi}}\,
\left(1-\frac{2\,\sigma}{\pi}\right)\Big].
\ea
This expression resums at NLO order the contributions 
near  the boundary point $\sigma={\pi\ov 2}$. 
This expression is not, of course,  expected to be 
correct near  $\sigma=0$, but it must reproduce the exact
 value of $\rho({\pi\ov 2})$ with  order $\oh{\eta}$ included. 
 This is true since
\be
\rho_{_{{\rm NLO}}}( {\pi\ov 2 }  ) =\left(\ln 2-\frac{1 }{2}\ln \eta  \right)+\frac{\eta }{4}+{\cal O}\left(\eta ^2\right)
\ee
is in agreement with the exact value of $\rho({\pi\ov 2 }  ) $ which is 
\be
\rho(   {\pi\ov 2 }          ) = {\rm arcsinh}\frac{1}{\sqrt\eta} = 
\left(\ln 2-\frac{1}{2}\ln \eta \right)+\frac{\eta }{4}-\frac{3 
\eta ^2}{32}+\frac{5 \eta ^3}{96}+{\cal O}\left(\eta ^4\right).
\ee
Also, as an additional check, we immediately
 reproduce  that $\rho'_{_{{\rm NLO}}}(    {\pi\ov 2 }        ) = 0$.

In order to obtain the resummation in a systematic way and to show that it 
comes from the behaviour of the string profile around the turning point
we can work out the expansion of the differential equation for $\rho(\sigma)$ around $\sigma = {\pi\ov 2 } $.
To this aim, let us  define 
\be
x = \kappa\left(\frac{\pi}{2}-\sigma\right),\qquad 
\quad \qr(x)\equiv \rho\left(\frac{\pi}{2}-\frac{x}{\kappa}\right),
\ee
and solve the  corresponding equation (cf. \rf{diff})
\ba
\qr'(x) =-\sqrt{1-\eta\,\sinh^2 \qr(x)}\ ,\  \ \ \ \ \ \ \ \ \ \ \ 
\qr(0) = \mbox{arcsinh}\frac{1}{\sqrt{\eta}}.
\ea
perturbatively in $\eta$, i.e. 
\ba
\qr(x) &=& \mbox{arcsinh}\frac{1}{\sqrt{\eta }}+\ln \mbox{sech}\, x-\frac{\eta}{4} \,x\, \tanh x \no \\
&& + \frac{\eta^2}{128}\,\left(
-1+\cosh(2x)-4\,x^2\,\mbox{sech}^2 x+10\,x\,\tanh x\right) + \oh{\eta^3} 
\ea
Here the  value $\qr(0)$ was  left unexpanded. Expanding it consistently we get  the final result
\ba
\qr(x) &=& 
\left(\frac{\pi  \kappa _0}{2} + \ln \frac{\text{sech}\,x}{2}\right)+\left(\frac{1}{4}
-\frac{1}{4} x \tanh x\right) \eta \no \\
&& + \left(-\frac{1}{32} x^2
   \text{sech}^2 x+\frac{1}{128} \cosh (2 x)+\frac{5}{64} \,x \,\tanh x-\frac{13}{128}\right) \eta ^2+\oh{\eta^3}
\ea
If we now use  the definition of $x =\kappa\left(\frac{\pi}{2}-\sigma\right)$ 
in this expression, we get precisely the NLO resummation in  \refeq{resummation} plus a new
$\eta^2$ term which is beyond the order of accuracy of \rf{eq:resummation}.

\

One may wonder  if  this systematic resummation of $\rho(\sigma)$ can be used to resum
 the associated contributions in the 1-loop  correction to string energy 
 discussed in section 3. This is not, however, immediately clear. 
 Plugging the expansion of $\rho$ around $\sigma=\frac{\pi}{2}$ in the 
$Q_\omega$ operator in \rf{part} 
 and denoting the resulting terms with  label ``fold'' to indicate the 
  expansion point, we find 
\be
Q_\omega = Q_{\omega, \rm fold}^{(0)} + \eta\,Q_{\omega, \rm fold}^{(1)} + \cdots,
\ee
where
\be
Q_{\omega, \rm fold}^{(0)} = 
\left(
\begin{array}{lll}
 -n^2-\omega ^2 &  - V_1  &  - V_2 \\
V_1 & n^2 +  \omega ^2 -  { 2 \ka \ov \omega } V_2   & -2 \omega  \kappa _0- V_2  \\
 V_2  & 2 \omega  \kappa _0+ V_2  &   n^2+\omega ^2-   { 2 \ka \ov \omega } V_2    
\end{array}  \right), \ee
\be  
z \equiv  \frac{\pi}{2}-\sigma\ , \ \ \ \ \ \ \ 
V_1 = \frac{\kappa _0^2}{\cosh^2\left(\kappa_0\,z\right)}+  2 i n \kappa _0 \tanh \left(\kappa_0\,z\right)  
\ , \ \ \ \  \ 
V_2 = \frac{\omega  \kappa_0}{\cosh^2 \left(\kappa_0\,z\right)} \ . 
\ea
In the large $\kappa_0$ limit, we can make the following replacements ($\int_0^\infty dz\,\delta_+(z) = 1$)
\ba
\kappa_0\,{\rm sech}^2(\kappa_0\,z) \longrightarrow \delta_+(z) \ , \ \ \ \ \ \ \ \ \ \ 
\kappa_0\,\tanh(\kappa_0\,z) \longrightarrow \kappa_0-\ln(2)\,\delta_+(z)\ .
\ea
After this substitution we can write
\be
Q_{\omega, \rm fold}^{(0)} &=& Q_\omega^{(0)} + Q^{(0)\ '}_{\omega, {\rm fold}}\,\delta_+(z), 
\ee
where $Q_\omega^{(0)}$ is the same operator \rf{awl}
we found in section 3 
in the expansion valid near  $\sigma = 0$,  while the new piece is
\be
Q^{(0)\ '}_{\omega, \rm fold} = 
\left(
\begin{array}{lll}
 0 & 2\, i\, n\, \ln 2-\kappa _0 & -\omega  \\
 -2\, i\, n\, \ln 2-\kappa _0 & -2 \kappa _0 & -\omega  \\
 \omega  & \omega  & -2 \kappa _0
\end{array}
\right)\ . 
\ee
This is a $\oh{\kappa_0}$ perturbation  over 
$Q_\omega^{(0)}$ whose matrix elements are  $\oh{\kappa_0^2}$
(since $n,\omega\sim\kappa_0$ in the combined sum and integral
like in \rf{met}). 
Unfortunately, 
higher order terms coming from 
this term can be estimated to have the same order of magnitude and thus must
 be resummed. In principle, the contribution from $Q_{\omega, \rm fold}^{(0)}$
must be treated exactly and separately, a task which we leave  for the future. 

Still, it   is encouraging  to note
that a possible non-zero contribution from the  near-turning-point region 
is expected to change the one-loop energy by a term proportional to 
\be
\frac{\kappa_0}{\kappa} = 1+\Big(\frac{1}{4}-\frac{1}{2\,\pi\,\kappa_0}\Big)\,\eta + \cdots  = 
1+\frac{1}{2\,\pi\,{\cal S}} +\cdots~.
\ee
This means that the induced modification of the coefficients in 
appearing in \rf{oka} 
must obey 
\be
\delta b_0 = \delta b_{11} = 0\ ,\ \ \  \qquad \delta b_{10} = \frac{1}{2\,\pi}\,\delta b_c\ .
\ee
Remarkably, this is precisely what is required by the reciprocity conditions in \rf{www},\rf{wer}.


\appendix
\subsection*{Appendix D:  Details of  large spin expansion for  folded  $(S,J)$
spinning string
}
\refstepcounter{section}
\def\theequation{D.\arabic{equation}}
\setcounter{equation}{0}


In this section we collect some details on large spin expansions used in  Section 2.2.

In the ``slow long strings'' regime ($\S \gg 1$,  $\J\ll \S$), the small $\eta$ expansions for
the ``anomalous'' part of the energy
and  the conformal spin read~\footnote{The expansions are obtained from (\ref{foldinform1}) and (\ref{foldinform2}) after the redefinition $\eta\rightarrow -1+16\eta+\sqrt{1+256\,\eta^2}$.}
\bea\no
\tg_{_{\J\ll1}}&=&\k+\frac{\k}{\o}\,\S-\S-\J\approx
\left[-\frac{1+\ln\eta}{\pi}+\frac{4(\ln\eta+12)}{\pi}\eta^2+{\cal O}(\eta^4)\right]-\J
\\\nonumber&&+\ \pi\,\J^2\left[
\frac{(1-\ln\eta)}{2\ln^2\eta}-\eta^2\left(\frac{10}{\ln\eta}+\frac{20}{\ln^2\eta}-\frac{44}{\ln^3\eta}
\right)+{\cal O}(\eta^4)
\right]+ ...\ , \no
\\
{\tilde s}_{_{\J\ll1}}&=&\S+\ha {\J} +\ha  {\tg} \approx
\left[\frac{1}{8\pi\eta}+\eta\frac{2\ln\eta+11}{2\pi}+{\cal O}(\eta^3)
\right]\no  \\
&&+\ \pi\, \J^2\left[\frac{1}{16\,\eta\ln^2\eta}-\eta\left(\frac{3}{2\ln\eta}-
\frac{13}{4\ln^2\eta}-\frac{11}{2\ln^3\eta}\right)+
{\cal O}(\eta^3)
\right]+...\ .
\eea
For the ``fast long strings'' ($\S\gg 1 $, with $\ln\S\ll\J\ll\S$) one finds
\bea\nonumber
&&\tg_{_{\ln\S\ll\J\ll\S}}\approx
\frac{1}{\pi^2\,\J}\left[\ln\eta\left(1+\frac{1}{2}\ln\eta\right)+44\,\eta^2\left(\frac{1}{11}
\ln^2\eta-\ln\eta-1\right)+{\cal O}(\eta^4)
\right]
\\\nonumber
&&+\frac{1}{\pi^4\J^3}\left[-\frac{1}{8}\left(\ln^4\eta+4\,\ln^3\eta\right)+2\,\eta^2\,
(-5\ln^4\eta+5\ln^3\eta
+33\ln^2\eta)+{\cal O}(\eta^4)\right]+...\ , \\
&&{\tilde s}_{_{\ln\S\ll\J\ll\S}}\approx\J\left[-\frac{1}{8\,
\eta\,\ln\eta}+\eta\left(1-\frac{11+12\ln\eta}{2\ln^2\eta}\right)+{\cal O}(\eta^3)
\right]\nonumber \\
&&\!\!\!\!+\frac{1}{\pi^2\J^2}\left[-\frac{\ln\eta}{16\,\eta}-
\eta\left(\frac{3\,\ln^2\eta}{2}+4\ln\eta
-\frac{11}{4}\right)
+{\cal O}(\eta^3)\right]+...\ .
\eea
Since  the function $\tP= {{\rm f } \ov \sql}$ in \rf{ko} coincides with
 the anomalous dimension evaluated at   zero of the denominator in \rf{LB2}
  in both cases we get  an equation expressing the parameter
 $\eta$  in terms of only the odd powers of the Casimir $\C$.
  From the power series expressions for $\tg$,  even in $\S$, it
  follows that the function $\tP$  has expansion in
   even negative powers of the semiclassical Casimir $\C$.

Explicitly, the first few corrections  read,   for slow long strings
\bea
\tP&\approx&\Big[  \frac{\ln8\pi\C-1}{\pi}+\frac{\ln 8\pi\C +
1 }{16\,\pi^3\C^2} +{\cal O}(1/\C^4)\Big] -\J
\\\nonumber
&&+\pi\,\J^2\big[\frac{1}{2\ln8\pi\C}-\frac{3}{32\pi^2\,\C^2\ \ln8\pi\C}+{\cal O}(1/\C^4)\big]+... \ ,
\ee
where $\C=\S+\ha \J$ and dots indicate corrections in $\J$.
For fast long strings,
\bea
&&\tP\approx\frac{1}{\pi^2\,\J}\Big[\frac{\ln^2\hC}{2}-\ln\hC+
\frac{1}{16\,\hC^2}\Big(4\ln\hC+3+\frac{3}{\ln\hC}+\frac{7}{\ln^2\hC}+...
 \Big)+{\cal O}(1/\hC^4)
\Big]\no
\\\nonumber
&&\ \ \ \ \ \ \ \ \ \ \ \  \ \ \
- \ \frac{1}{\,\pi^4\,\J^3}\Big[\frac{\ln^4 \hC}{8}+\frac{1}{32\,\hC^2}
\Big(4\ln^3 \hC+5\ln^2\hC+9\ln\hC \no \\
&& \ \ \ \ \ \ \ \ \ \ \ \ \ \ \ \ +\ 16+\frac{24}{\ln\hC}+\frac{34}{\ln^2{\hC}}+...\Big)
+{\cal O}(1/\hC^4)\Big]+...\ ,
\ee
where $\hC={\C\ov \J}={\S\ov \J}+ \ha $ and dots inside
 round brackets indicate corrections in $1/\ln\hC$.
 As was already noted  in section 2.2,
  the expansion in the case  of the fast  long strings
 is not of the same type as in  \rf{qw}  and \rf{rec} assumed in the main part
 of this paper.

 \

Let us mention also that in the case of the $m$-folded  string
the interval $0\leq\s<2\pi$ is split into $4m$ segments: for $0<\s<\frac{\pi}{2m}$ the function
$\r(\s)$ increases reaching its maximal value $\rho_0$,
then decreases to zero for $\frac{\pi}{2m}\leq\s\leq\frac{\pi}{m}$, etc.
This implies the condition
\be
2\pi=\int_0^{2\pi}d\s=4\,m\,\int_0^{\rho_0}
\frac{d\r}{\sqrt{(\k^2-\J^2)\cosh^2\r-(\o^2-\n^2)\sinh^2\r}} \ ,
\ee
which leads to  a
factor of $m$ in front of the relevant expressions for
$\E$, $\S$, $\sqrt{\k^2-\J^2}$. The large spin expansion is then similar to the $m=1$ case.
Once $\E$ is expressed in terms of $\S$ and $\J$, the parameter
$m$ enters only  in combination with the string
tension $\sql \ov 2\pi$.

\appendix
\refstepcounter{section}
\def\theequation{E.\arabic{equation}}
\setcounter{equation}{0}

\subsection*{Appendix E:  Higher order relations  from reciprocity at  strong coupling }

The evidence for the functional relation  and reciprocity \rf{ko},\rf{rec} at weak coupling
suggests that  the corresponding constraints should hold  also in strong-coupling expansion.
As we have seen, the   large spin expansion of  anomalous dimensions  at strong coupling
appears to  have the same structure as at weak coupling \rf{xa} where now
\be
&& f \equiv  \sql\ \bar f\ , \ \ \ \ \ \ \   \bar f =  a_0+\frac{b_0}{\sqrt{\l}}+\frac{c_0}{(\sqrt{\l})^2}+...
 \ , \la{fe}\\
&& f_c\equiv  \sql\ \bar f_c\ , \ \ \ \ \ \ \
 \bar f_c= a_c+\frac{b_c}{\sqrt{\l}}+\frac{c_c}{(\sqrt{\l})^2}+... \ , \la{fee}\\
&&
f_{mk} \equiv  (\sql)^{m+1} \ \bar f_{mk}\ , \ \ \ \ \ \ \ \
 \bar f_{mk}= a_{mk}+\frac{b_{mk}}{\sqrt{\l}}+\frac{c_{mk}}{(\sqrt{\l})^2}+... \ . \la{ef}
 \ee
Assuming the functional relation or \rf{lea},  one is then able to compute the coefficients
 $f_{mm}$
of  $\ln^m S \ov S^m$    in terms of the strong-coupling expansion coefficients
in the scaling function $f$. The latter are  known up to 2-loop order
directly from the string-theory computations
\ci{ft1,rt}
\be
a_0= \frac{1}{\pi}\ , \ \ \ \ \ \
b_0= -\frac{1}{\pi} \ {3\ln 2} \ , \ \ \ \ \ \
c_0= -\frac{1}{\pi} {\rm K} \ , \ \ \ .... \la{res} \ee
and also to a high (in principle,  arbitrarily high)  order
 from the analytic strong coupling
solution \ci{bkk} of   the BES \ci{bes} equation  for the function $f$.
This means that $f_{mm}$ are then effectively determined
if  the functional relation applies.

Assuming the validity of the
  reciprocity condition \rf{rec}  should lead to additional constraints on the
subleading  coefficients
like \rf{VV} which here should be understood in terms of  power series in $1 \ov \sql$.
As a result, one should find non-trivial relations between strong-coupling
expansion coefficients in \rf{fee} and \rf{ef}.

There is, however, a subtlety in  formulating  the reciprocity condition in the context of
large spin expansion at strong coupling
 as defined by string
 semiclassical perturbation theory   where   all non-zero charges are
automatically large at large $\l$.
 For example,  the case of finite twist $J=2,3,...$
can not be distinguished from the formal case of $J=0$.
It is usually assumed
that the folded string in $AdS_5$ with zero angular momentum in $S^5$
describes an operator  of small twist, but that can be
 $J=2$ or $J=3$, etc. To establish a relation to the definition of reciprocity
 in weakly coupled gauge theory expansion with finite twist
 one would need  to consider the case
 of semiclassical $(S,J)$ string and then resum the series  for its energy
 (both in $J$ and in $\sqrt \l$)
 so that the limit  of finite $J$  would  make sense.

 Here
 we shall assume that in  checking the reciprocity \rf{rec}  at
 subleading order in strong coupling at $J=0$
 one may  simply take the Casimir $C$ in \rf{rec} as $C= \sql \,{\cal C}, \ {\cal  C }
 = \S$,
 and ignore the shifts in brackets in \rf{MVV} or \rf{VV}, i.e. getting
 \bea\la{VVs}
\bar f_{10} ={1 \ov 2}  \bar f  \bar f_c  \ , \ \ \ \ \ \ \
\bar f_{32} = {\frac{1}{16}} \bar f \ (\bar  f^3 - 2\bar  f^2\  \bar f_c-  16\bar f_{21}) \ ,
 \ \ ... \ .
\eea
Equivalently, these relations follow from \rf{MVV} 
by noting that at strong coupling $f \sim f_c \sim \sqrt{\lambda}$   and thus 
terms of order 1 or  $J \ll sqrt{\lambda}$  can be ignored. 

Multiplying the series in \rf{fe}--\rf{ef}
we then  find that some of the  1-loop coefficients
can be  expressed in terms of the  tree-level
coefficients
and the coefficients in $f$. Explicitly,
\be
b_{11}&=&a_0 b_0 ~,   ~~~~~~~~b_{10}={ 1 \ov 2}  ( a_0 b_c + a_c \,b_0)
~, ~~~~~~~~
b_{22}=-\frac{3}{8}a_0^2\,b_0    \, \la{amm} \\
b_{33}&=& {1 \ov 6}  a_0^3\,b_0 \ , ~~~~~~
b_{32}={1 \ov 8} a_0^3 (  2 b_0 - b_c)   -a_{21}\,b_0-\frac{3}{8}a_0^2\,a_c\,b_0-a_0\,b_{21}  \  , ...
  \label{aml}
\eea
We have verified the  validity  of these relations for $b_{11}$
 and $b_{10}$   in section \ref{sec:1loop}.


\appendix
\refstepcounter{section}
\def\theequation{F.\arabic{equation}}
\setcounter{equation}{0}

\subsection*{Appendix F:  Large $S$ expansions for twist 2 and twist 3 anomalous dimensions
 at weak coupling}
\label{app:gaugexp}

Here we shall  collect  the coefficients of
large spin  expansion
of anomalous dimensions for planar SYM operators of twist 2 and 3,
up to four loops in the gauge coupling and up to $1/S^3$ order.~\footnote{In
 the case of twist 3 operators, the anomalous dimensions we will consider here are the
 \emph{minimal} in the band.}
   They are derived from the closed expressions in terms of the harmonic
  sums that were obtained (mainly exploiting the maximum transcendentality principle and the
  asymptotic Bethe ansatz), respectively,  in
  ~\cite{Kotikov:2007cy}  (at four loops) for
 the twist two scalar sector,
  in~\cite{Beccaria:2007cn,Kotikov:2007cy} for the
  twist three scalar sector and in~\cite{Beccaria:2007pb} (at three loops) and~\cite{bf}
   (at four loops) for the ``gauge'' sector.

All these expansions were proven to satisfy the reciprocity
property, for a review
  see~\cite{bf}. The expansions are indeed of the generic form (\ref{xa}) where the
   coefficients satisfy the relations (\ref{MVV}) once $J$ (and the flavor index
    $\ell$, see footnote 9 in the Introduction)
   are fixed  accordingly.

The coefficients of the  leading  $\ln^m{S}/S^m$  terms below  are manifestly
 universal in twist and flavor.\foot{The only exception being the four loop coefficient of the term $
 \ln^2 S/S^2$ in  the case of twist two scalar operators.
  However,
  it seems  reasonable to relate this exception to the wrapping-induced breakdown of the Bethe equations
   at four loops for twist two operators.}
     As far as these leading terms are concerned, there is no need to
   explicitly write down  the results for the   twist two gaugino and gauge sectors, and
    the twist
   three gaugino sector. Indeed, the closed formulas for their anomalous dimensions can be
   deduced from  the one for  the twist two scalar case by just shifting the argument of the harmonic
   sums\footnote{It is well known that in $\mathcal{N}=4$ SYM all twist two operators belong to
    the same supermultiplet, and their anomalous dimension is expressed in terms of
    a universal function  with shifted arguments
$
\g^\varphi_{J=2}(S)=\g_{{\rm univ}}(S)~,~~\g^\psi_{J=2}(S)=\g_{{\rm univ}}(S+1)~,~~\g^A_{
J=2}(S)=\g_{{\rm univ}}(S+2).
$
In~\cite{Beccaria:2007vh} it was proved that  the anomalous dimension for twist three operators
 built out of  gauginos  is related to the one of the twist two universal supermultiplet as
$
\gamma^\psi_{J=3}(S) = \gamma^\varphi_{J=2}(S+2).
$
}
 but such shifts  do not affect   the coefficients of  the leading  $\ln^m S/S^m$ terms.
It is worth stressing again  that this universality, a well-known feature of
 the leading $\ln S$ coefficient  (or cusp anomaly),
  is a nontrivial consequence of the functional relation (\ref{ko}),
  as was noticed in~\cite{Beccaria:2007pb} and emphasized in~\cite{bkp}.

At weak coupling it  is useful to rewrite \rf{xa} as
\bea\nonumber
\gamma(S)_{_{S \gg1 }}  &=& f\,\ln\,\bS
+ \bar f_c + \frac{f_{11}\,\ln\,\bS+\bar f_{10}}{S} + \frac{f_{22}\,\ln^2\,\bS+\bar f_{21}\,\ln\,\bS+\bar
f_{20}}{S^2}+\\
&& \ \ \ \ \ \ \ \ \ +\  \frac{f_{33}\,\ln^3\,\bS +\bar  f_{32}\,\ln^2\,\bS+\bar f_{31}\,\ln\,\bS+
\bar f_{30}}{S^3}+{\cal O}\big({\ln^4\,\bS\ov S^4}\big),  \la{xak}
\eea
where $\bar S=e^{\g_E}S$ and  the coefficients will be power series in  $\hat\l=\frac{\l}{16\pi^2}$.
Then one finds:

\

\noindent
{{\bf Twist two scalar sector:}}
\bea\nonumber
&&f=8 \hat\l-\frac{8 \pi ^2 }{3}\hat\l^2+\frac{88
   \pi ^4 }{45}\hat\l^3- (\frac{584 \pi ^6}{315}+ 64 \zeta_3^2) \hat\l^4 \ , \no \\
&&\bar f_c=    -24 \zeta_3 \hat\l ^2+(\frac{16}{3} \pi ^2 \zeta_3+160
   \zeta_5) \hat\l ^3+  (-\frac{56}{15} \pi ^4 \zeta_3-\frac{80}{3} \pi ^2 \zeta_5-1400
   \zeta_7) \hat\l ^4 \, \no \\
&&f_{11}=  32\hat\l^2-\frac{64\,\pi^2}{3}\hat\l^3+\frac{96\,\pi^4}{5}\hat\l^4\ , \no \\
&&\bar f_{10}=  4 \hat\l-\frac{4 \pi ^2}{3} \hat\l ^2+  (\frac{44 \pi ^4}{45}-96
   \zeta_3) \hat\l ^3+ (-\frac{292 \pi ^6}{315}+\frac{160}{3} \pi ^2 \zeta_3-32 \zeta_3^2
   +640 \zeta_5) \hat\l ^4   \ , \no \\
&&f_{22}=    -64\hat\l^3+(64\pi^2-128\zeta_3)\hat\l^4 \ , \\
 &&\bar f_{21}=   -16\hat\l^2+(128+\frac{16\,\pi^2}{3})\hat\l^3+(-128 \pi^2 - \frac{32\,\pi^4}
   {15}+ 448 \zeta_3)\hat\l^4 \ , \no \\
  &&\bar f_{20}=   -\frac{2}{3}\hat\l+ (24+\frac{2 \pi
   ^2}{9} ) \hat\l ^2- (\frac{32 \pi ^2}{3}+\frac{22 \pi
   ^4}{135}-48 \zeta_3  )\hat \l ^3 \no \\
  &&  \ \ \ \ \ \ \ + (\frac{136 \pi ^4}{15}+\frac{146 \pi ^6}{945}-
   384 \zeta_3-\frac{32}{3}
    \pi ^2 \zeta_3
     +\frac{16 \zeta_3^2}{3}-320 \zeta_5 ) \hat\l ^4 \no \\
 &&f_{33}=  \frac{512}{3}\,\hat\l^4  \ ,\no \\
  &&\bar f_{32}=  64\,\hat\l^3+( -768 - \frac{64 \,\pi^2}{3} + 128 \zeta_3)\hat\l^4\ , \no \\
  &&\bar f_{31 }=
   \frac{16
  }{3} \hat\l^2+ (-256+\frac{16 \pi ^2}{9} ) \hat\l^3+ (512+\frac{512 \pi ^2}{3}-\frac{64
   \pi ^4}{15}-576 \zeta_3 ) \hat \l ^4\ , \no \\
 &&\bar f_{30 }=   -\frac{56}{3}\hat \lambda ^2+(96+\frac{40 \pi ^2}{9}-16 \zeta_3 )\hat \lambda ^3
-(\frac{224 \pi ^2}{3}+\frac{32 \pi ^4}{15}-800 \zeta_3+\frac{64}{9} \pi ^2 \zeta_3 -\frac{320
 \zeta_5}{3}) \hat \lambda ^4   \no
\eea
\noindent
{{\bf Twist three scalar sector:}}
\bea\nonumber
&&f=8 \hat\l-\frac{8 \pi ^2 }{3}\hat\l^2+\frac{88
   \pi ^4 }{45}\hat\l^3- (\frac{584 \pi ^6}{315}+ 64 \zeta_3^2) \hat\l^4 \ , \no \\
&&\bar f_c= -8 \ln2 \hat\lambda +(\frac{8}{3} \pi ^2 \ln 2-8 \zeta_3 ) \hat\lambda
   ^2+(-\frac{88}{45} \pi ^4 \ln2+\frac{8}{3} \pi ^2 \zeta_3-8 \zeta_5 )
\hat   \lambda ^3 \no \\
&&\ \ \ \ \ \ \ \ \ +\frac{8}{315} (73 \pi ^6 \ln2-84 \pi ^4 \zeta_3+2520 \ln2 \zeta_3^2+105 \pi ^2 \zeta_5+
17325 \zeta_7)\hat \lambda ^4, \no \\
&&f_{11}=  32\hat\l^2-\frac{64\,\pi^2}{3}\hat\l^3+\frac{96\,\pi^4}{5}\hat\l^4\ , \no \\
&&\bar f_{10}= 8 \hat\lambda +(-\frac{8 \pi ^2}{3}-32 \ln2) \hat\lambda ^2+(\frac{88 \pi
   ^4}{45}+\frac{64}{3} \pi ^2 \ln2-32 \zeta_3 ) \hat\lambda ^3\no \\
 &&\ \ \ \ \ \ \ \ \  -\frac{8}{315}(73
   \pi ^6+756 \pi ^4 \ln2-840 \pi ^2 \zeta_3+2520 \zeta_3^2+1260 \zeta_5 )
  \hat \lambda ^4 , \no \\
&&f_{22}=    -64\hat\l^3+ 64\pi^2 \,\hat\l^4 \ ,\no \\
 &&\bar f_{21}=   -32 \hat\lambda ^2+(128+\frac{64 \pi ^2}{3}+128 \ln2) \hat\lambda ^3+(-256-
  128 \pi^2-\frac{96 \pi ^4}{5}-128 \pi ^2 \ln2+256 \zeta_3) \hat\lambda ^4,\no
  \eea
  \bea
&&\bar f_{20}=   -\frac{8 \lambda }{3}+(48+\frac{8 \pi ^2}{9}+32 \ln2)\hat \lambda
   ^2 \no \\
   &&\ \ \ \ \ \ \ \ +(32-\frac{80 \pi ^2}{3}-\frac{88 \pi ^4}{135}-128 \ln2-\frac{64}{3} \pi ^2 \ln2
    -64 \ln^22+32 \zeta_3) \hat\lambda ^3 \no \\
    && \ \ \ \ \ \ \ \ \ +(-512-\frac{32 \pi ^2}{3}+\frac{352
   \pi ^4}{15}+\frac{584 \pi ^6}{945}+256 \ln2+128 \pi ^2\ln2
   +\frac{96}{5} \pi ^4\ln2\no \\
  &&\ \ \ \ \ \ \ \ \   +64 \pi ^2 \ln^22-128 \zeta_3 -\frac{64}{3} \pi ^2 \zeta_3 -256 \ln2
   \zeta_3+\frac{64 \zeta_3^2}{3}+32 \zeta_5 ) \hat\lambda ^4,\no \\
 &&f_{33}=  \frac{512}{3}\,\hat\l^4  \ , \\
  &&\bar f_{32}=   128 \hat\lambda ^3-  (768+128 \pi ^2+512 \ln2) \hat\lambda ^4\ ,\no \\
  &&\bar f_{31}=  \frac{64}{3}\hat\l^2+(-512-\frac{128 \pi ^2}{9}-256 \ln2) \hat\lambda
   ^3\no \\
 &&\ \ \ \ \ \ \   +(768+\frac{1408 \pi ^2}{3}+\frac{64 \pi ^4}{5}+1536 \ln2
   +256 \pi ^2 \ln2+512 \ln^22-512 \zeta_3 )\hat \lambda ^4,\no \\
   &&\bar f_{30}=  -(\frac{224}{3}+\frac{64 \ln 2}{3}) \hat\lambda ^2+(128+\frac{352 \pi
   ^2}{9}+512 \ln2+\frac{128}{9} \pi ^2 \ln2+128 \ln^22-\frac{64 \zeta_3}{3})\hat \lambda ^3\no \\
 &&\ \ \ \ \ \ \   +(896-\frac{448 \pi ^2}{3}-\frac{512 \pi ^4}{15}-768\ln2
-\frac{1408}{3} \pi ^2 \ln2-\frac{64}{5} \pi ^4 \ln2-768 \ln^22\no \\
&&\ \ \  \ \ \ \  -128 \pi ^2
   \ln^22-\frac{512 \ln ^32}{3}+640 \zeta_3 +\frac{128}{9} \pi ^2 \zeta_3+512 \ln2
\zeta_3 -\frac{64 \zeta_5}{3}) \hat\lambda ^4.\no
\eea
\noindent
{{\bf Twist three ``gauge"  sector:}}
\bea\nonumber
&&f=8 \hat\l-\frac{8 \pi ^2 }{3}\hat\l^2+\frac{88
   \pi ^4 }{45}\hat\l^3- (\frac{584 \pi ^6}{315}+ 64 \zeta_3^2) \hat\l^4 \ , \no \\
&&\bar f_c=8(1- \ln2)\hat \lambda +\frac{8}{3} (-12-\pi ^2+\pi ^2 \ln2-3 \zeta_3)
   \hat\lambda ^2-\frac{8}{45} (-1440-60 \pi ^2-11 \pi ^4\no \\
   &&\ \ \ \  +11 \pi ^4 \ln2 -15 \pi ^2 \zeta_3+45 \zeta_5) \hat\lambda ^3+\frac{8}{315} (-100800-3360 \pi ^2
   -336 \pi ^4\no \\
    &&\ \ \ \
   -73
   \pi ^6+73 \pi ^6 \ln2-84 \pi ^4 \zeta_3
   -2520 \zeta_3^2+2520\ln2 \zeta_3^2+105 \pi ^2 \zeta_5+17325 \zeta_7) \hat\lambda ^4    \no \\
&&f_{11}=  32\hat\l^2-\frac{64\,\pi^2}{3}\hat\l^3+\frac{96\,\pi^4}{5}\hat\l^4\ , \no \\
&&\bar f_{10}=32 \hat\lambda +(32-\frac{32 \pi ^2}{3}-32 \ln2)  \hat\lambda ^2+\frac{32}{45}
   (-180-30 \pi ^2+11 \pi ^4+30 \pi ^2 \ln2-45 \zeta_3) \hat \lambda
   ^3\no \\\nonumber
   &&\ \ \ \ -\frac{32}{315} (-10080-840 \pi ^2-189 \pi ^4+73 \pi ^6+189 \pi ^4 \ln2-210 \pi
   ^2 \zeta_3 +2520 \zeta_3^2+315 \zeta_5) \hat \lambda ^4\\\nonumber
&&f_{22}=    -64\hat\l^3+64\pi^2\,\hat\l^4 \ , \\
   &&\bar f_{21}= -128 \hat\lambda ^2+(\frac{256 \pi ^2}{3}+128 \ln2)\hat \lambda ^3+(256
-\frac{384
   \pi ^4}{5}-128 \pi ^2 \ln2)\hat \lambda ^4\no \\\nonumber
&&\bar f_{20}=-\frac{200 }{3}\hat\l+(16+\frac{200 \pi ^2}{9}+128 \ln2)\hat \lambda
   ^2+(480-\frac{16 \pi ^2}{3}-\frac{440 \pi ^4}{27}-\frac{256}{3} \pi ^2 \ln2\no\\
  &&\ \  \ \  -64
   \ln^22+128 \zeta_3)\hat \lambda ^3+(-2816-\frac{1120 \pi ^2}{3}+\frac{64 \pi
   ^4}{15}+\frac{2920 \pi ^6}{189}-256 \ln2\no \\
  &&\ \   +\frac{384}{5} \pi ^4 \ln2+64 \pi ^2 \ln^22
  -128 \zeta_3-\frac{256}{3} \pi ^2 \zeta_3+\frac{1600 \zeta_3^2}{3}+128 \zeta_5) \hat\lambda ^4\no
   \eea
   \bea
   &&f_{33}=  \frac{512}{3}\,\hat\l^4  \ , \\
 &&\bar f_{32}=512 \hat\lambda ^3+(-256-512 \pi ^2-512 \ln2)\hat \lambda ^4\no \\\nonumber
 &&\bar f_{31}=\frac{1600 }{3}\hat\lambda ^2+(-640-\frac{3200 \pi ^2}{9}-1024 \ln2) \hat\lambda
   ^3\no \\
 &&\ \  \ \    +(-1792+\frac{1792 \pi ^2}{3}+320 \pi ^4+512 \ln2
 +1024 \pi ^2 \ln2+512 \ln^22)\hat \lambda ^4 \no \\
&&\bar f_{30}=192 \hat\lambda +(-\frac{1120}{3}-64 \pi ^2-\frac{1600 \ln2}{3}) \hat\lambda
   ^2+(-\frac{5824}{3}+\frac{1856 \pi ^2}{9}+\frac{704 \pi ^4}{15}+640\ln2\no \\
&&\ \ \ \ \no   +\frac{3200}{9} \pi ^2 \ln2+512 \ln^22-\frac{1600 \zeta_3}{3})\hat \lambda
   ^3+(\frac{25984}{3}+\frac{15488 \pi ^2}{9}-\frac{544 \pi ^4}{3}-\frac{4672 \pi
   ^6}{105}\\\nonumber
   &&\ \ \ \no
+1792 \ln2-\frac{1792}{3} \pi ^2 \ln2-320 \pi ^4\ln2-256 \ln^22
 -512 \pi ^2 \ln^22-\frac{512 \ln ^32}{3}+1152 \zeta_3 \no \\
   && \ \ \ \
+\frac{3200}{9} \pi ^2
   \zeta_3
-1536 \zeta_3^2-\frac{1600 \zeta_5}{3})\hat \lambda ^4\no
\eea


\newpage

\baselineskip 9pt

\begin{thebibliography}{99}

\bi{bes}
N.~Beisert, B.~Eden and M.~Staudacher,
  ``Transcendentality and crossing,''
  J.\ Stat.\ Mech.\  {\bf 0701}, P021 (2007)
  [arXiv:hep-th/0610251].


\bi{frs}
  L.~Freyhult, A.~Rej and M.~Staudacher,
  ``A Generalized Scaling Function for AdS/CFT,''
  arXiv:0712.2743 [hep-th].




 \bi{bgk}
  A.~V.~Belitsky, A.~S.~Gorsky and G.~P.~Korchemsky,
  { Logarithmic scaling in gauge / string correspondence},
  Nucl.\ Phys.\  B {\bf 748}, 24 (2006)
  [arXiv:hep-th/0601112].



\bi{bkp}
  A.~V.~Belitsky, G.~P.~Korchemsky and R.~S.~Pasechnik,
 ``Fine structure of anomalous dimensions in N=4 super Yang-Mills theory,''
  arXiv:0806.3657 [hep-ph].




\bi{korch}
G.~P.~Korchemsky,
  ``Quasiclassical QCD pomeron,''
  Nucl.\ Phys.\  B {\bf 462}, 333 (1996)
  [arXiv:hep-th/9508025].


\bibitem{gkp}
  S.~S.~Gubser, I.~R.~Klebanov and A.~M.~Polyakov,
  ``A semi-classical limit of the gauge/string correspondence,''
  Nucl.\ Phys.\ B {\bf 636}, 99 (2002)
  [hep-th/0204051].


\bi{dev}
H.~J.~de Vega and I.~L.~Egusquiza,
  ``Planetoid String Solutions in 3 + 1 Axisymmetric Spacetimes,''
  Phys.\ Rev.\ D {\bf 54}, 7513 (1996)
  [hep-th/9607056].

\bi{pt}
I.-Y. Park and A.A. Tseytlin, unpublished (2005).

\bi{tt}
A.~Tirziu and A.~A.~Tseytlin,
  ``Quantum corrections to energy of short spinning string in AdS5,''
  arXiv:0806.4758 [hep-th].




\bibitem{ft1}
  S.~Frolov and A.~A.~Tseytlin,
  ``Semiclassical quantization of rotating superstring in AdS(5) x S(5),''
  JHEP {\bf 0206}, 007 (2002)
  [arXiv:hep-th/0204226].



\bibitem{kot}
  A.~V.~Kotikov, L.~N.~Lipatov, A.~I.~Onishchenko and V.~N.~Velizhanin,
  ``Three-loop universal anomalous dimension of the Wilson operators in N =  4
  SUSY Yang-Mills model,''
  Phys.\ Lett.\  B {\bf 595}, 521 (2004)
  [Erratum-ibid.\  B {\bf 632}, 754 (2006)]
  [arXiv:hep-th/0404092].


 \bi{moch}
  S.~Moch, J.~A.~M.~Vermaseren and A.~Vogt,
  {``The three-loop splitting functions in QCD: The non-singlet case''},
  Nucl.\ Phys.\  B {\bf 688}, 101 (2004)
  [arXiv:hep-ph/0403192];
  A.~Vogt, S.~Moch and J.~A.~M.~Vermaseren,
  {``The three-loop splitting functions in QCD: The singlet case''},
  Nucl.\ Phys.\  B {\bf 691}, 129 (2004)
  [arXiv:hep-ph/0404111].



 \bi{dok1}
  Yu.~L.~Dokshitzer, G.~Marchesini and G.~P.~Salam,
 ``Revisiting parton evolution and the large-x limit,''
  Phys.\ Lett.\  B {\bf 634}, 504 (2006)
  [arXiv:hep-ph/0511302].

\bi{Alt}
  G.~Altarelli,
  ``Partons In Quantum Chromodynamics,''
  Phys.\ Rept.\  {\bf 81}, 1 (1982).




 \bibitem{bk}
  B.~Basso and G.~P.~Korchemsky,
  ``Anomalous dimensions of high-spin operators beyond the leading order,''
  Nucl.\ Phys.\  B {\bf 775}, 1 (2007)
  [arXiv:hep-th/0612247].
  G. Korchemsky, ``Anomalous dimensions  of high-spin operators   beyond the
  leading order'', talk at the 12th Claude Itzykson Meeting,  Saclay,
  June (2007).

 \bi{dok2}
  Yu.~L.~Dokshitzer and G.~Marchesini,
  ``N = 4 SUSY Yang-Mills: Three loops made simple(r),''
  Phys.\ Lett.\  B {\bf 646}, 189 (2007)
  [arXiv:hep-th/0612248].



\bibitem{Kotikov:2007cy}
  A.~V.~Kotikov, L.~N.~Lipatov, A.~Rej, M.~Staudacher and V.~N.~Velizhanin,
  ``Dressing and Wrapping,''
  J.\ Stat.\ Mech.\  {\bf 0710}, P10003 (2007)
  [arXiv:0704.3586 [hep-th]].

\bibitem{bf}
  M.~Beccaria and V.~Forini,
  ``Reciprocity of gauge operators in N=4 SYM,''
  arXiv:0803.3768



 \bi{kor88}
 G.~P.~Korchemsky,
  ``Asymptotics of the Altarelli-Parisi-Lipatov Evolution Kernels of Parton
  Distributions,''
  Mod.\ Phys.\ Lett.\  A {\bf 4}, 1257 (1989).


 \bi{ar}
 L.~F.~Alday and R.~Roiban,
  ``Scattering Amplitudes, Wilson Loops and the String/Gauge Theory
  Correspondence,''
  arXiv:0807.1889 [hep-th].


 \bi{kruz}
  M.~Kruczenski,
  ``A note on twist two operators in N = 4 SYM and Wilson loops in Minkowski
  signature,''
  JHEP {\bf 0212}, 024 (2002)
  [arXiv:hep-th/0210115].


 \bi{am}
 L.~F.~Alday and J.~M.~Maldacena,
  ``Gluon scattering amplitudes at strong coupling,''
  JHEP {\bf 0706}, 064 (2007)
  [arXiv:0705.0303 [hep-th]].



 \bi{krtt}
  M.~Kruczenski, R.~Roiban, A.~Tirziu and A.~A.~Tseytlin,
  ``Strong-coupling expansion of cusp anomaly and gluon amplitudes from quantum
  open strings in $AdS_5 x S^5$,''
  Nucl.\ Phys.\  B {\bf 791}, 93 (2008)
  [arXiv:0707.4254 [hep-th]].


 \bi{dix}
  L.~J.~Dixon, L.~Magnea and G.~Sterman,
  ``Universal structure of subleading infrared poles in gauge theory
  amplitudes,''
  JHEP {\bf 0808}, 022 (2008)
  [arXiv:0805.3515 [hep-ph]].

 \bi{alm}
 L.~F.~Alday, talk given at ``Strings 2008'', August 2008, CERN, Geneva.

\bibitem{bdm}
  M.~Beccaria, Yu.~L.~Dokshitzer and G.~Marchesini,
  ``Twist 3 of the sl(2) sector of N=4 SYM and reciprocity respecting
  evolution,''
  Phys.\ Lett.\  B {\bf 652}, 194 (2007)
  [arXiv:0705.2639 [hep-th]].


 \bi{dor}
 N.~Dorey,
  ``A Spin Chain from String Theory,''
  arXiv:0805.4387 [hep-th].

\bi{ggg}
V.~M.~Braun, S.~E.~Derkachov, G.~P.~Korchemsky and A.~N.~Manashov,
  ``Baryon distribution amplitudes in {QCD},''
  Nucl.\ Phys.\  B {\bf 553}, 355 (1999)
  [arXiv:hep-ph/9902375].
S.~E.~Derkachov, G.~P.~Korchemsky and A.~N.~Manashov,
  ``Evolution equations for quark gluon distributions in multi-color QCD  and
  open spin chains,''
  Nucl.\ Phys.\  B {\bf 566}, 203 (2000)
  [arXiv:hep-ph/9909539].

\bibitem{bc}
  M.~Beccaria and F.~Catino,
  ``Large spin expansion of the long-range Baxter equation in the sl(2) sector
  of N=4 SYM,''
  JHEP {\bf 0801}, 067 (2008)
  [arXiv:0710.1991 [hep-th]].




 \bi{bra}
  V.~M.~Braun, G.~P.~Korchemsky and D.~Mueller,
  ``The uses of conformal symmetry in QCD,''
  Prog.\ Part.\ Nucl.\ Phys.\  {\bf 51}, 311 (2003)
  [arXiv:hep-ph/0306057].

\bibitem{Laenen:2008ux}
  E.~Laenen, L.~Magnea and G.~Stavenga,
  ``On next-to-eikonal corrections to threshold resummation for the Drell-Yan
  and DIS cross sections,''
  arXiv:0807.4412 [hep-ph].


\bibitem{Beisert:2005fw}
  N.~Beisert and M.~Staudacher,
  ``Long-range $PSU(2,2|4)$ Bethe ansaetze for gauge theory and strings,''
  Nucl.\ Phys.\  B {\bf 727}, 1 (2005)
  [arXiv:hep-th/0504190].

\bibitem{Beccaria:2007pb}
  M.~Beccaria,
  ``Three loop anomalous dimensions of twist-3 gauge operators in N=4 SYM,''
  JHEP {\bf 0709}, 023 (2007)
  [arXiv:0707.1574 [hep-th]].

\bibitem{kru}
  M.~Kruczenski,
  ``Spiky strings and single trace operators in gauge theories,''
  JHEP {\bf 0508}, 014 (2005)
  [hep-th/0410226].


\bibitem{bfst}
  N.~Beisert, S.~Frolov , M.~Staudacher and A.~A.~Tseytlin,
  ``Precision spectroscopy of AdS/CFT,''
  JHEP {\bf 0310}, 037 (2003)
  [hep-th/0308117].


 \bi{ftt}
  S.~Frolov, A.~Tirziu and A.~A.~Tseytlin,
  ``Logarithmic corrections to higher twist scaling at strong coupling from
  AdS/CFT,''
  Nucl.\ Phys.\  B {\bf 766}, 232 (2007)
  [hep-th/0611269].



\bibitem{bkk}
  B.~Basso, G.~P.~Korchemsky and J.~Kotanski,
  ``Cusp anomalous dimension in maximally supersymmetric Yang-Mills theory at
  strong coupling,''
  Phys.\ Rev.\ Lett.\  {\bf 100}, 091601 (2008)
  [0708.3933 [hep-th]].

\bibitem{rtt}
  R.~Roiban, A.~Tirziu and A.~A.~Tseytlin,
  ``Two-loop world-sheet corrections in $AdS_5 x S^5$ superstring,''
  JHEP {\bf 0707}, 056 (2007)
  [0704.3638 [hep-th]]. revised (v4) 08/07

\bibitem{rt}
  R.~Roiban and A.~A.~Tseytlin,
  ``Strong-coupling expansion of cusp anomaly from quantum superstring,''
  JHEP {\bf 0711}, 016 (2007)
  [0709.0681 [hep-th]].

\bi{rt2}
R.~Roiban and A.~A.~Tseytlin,
  ``Spinning superstrings at two loops: strong-coupling corrections to
  dimensions of large-twist SYM operators,''
  Phys.\ Rev.\  D {\bf 77}, 066006 (2008)
  [arXiv:0712.2479 [hep-th]].

\bi{kri}
P.~Y.~Casteill and C.~Kristjansen,
  ``The Strong Coupling Limit of the Scaling Function from the Quantum   String
  Bethe Ansatz,''
  Nucl.\ Phys.\  B {\bf 785}, 1 (2007)
  [arXiv:0705.0890 [hep-th]].



\bi{grm}
N.~Gromov,
  ``Generalized Scaling Function at Strong Coupling,''
  JHEP {\bf 0811}, 085 (2008)
  [arXiv:0805.4615 [hep-th]].





\bibitem{dgt}
  N.~Drukker, D.~J.~Gross and A.~A.~Tseytlin,
  ``Green-Schwarz string in AdS(5) x S(5): Semiclassical partition  function,''
  JHEP {\bf 0004}, 021 (2000)
  [arXiv:hep-th/0001204].



\bi{sak}
S.~Schafer-Nameki and M.~Zamaklar,
  ``Stringy sums and corrections to the quantum string Bethe ansatz,''
  JHEP {\bf 0510}, 044 (2005)
  [arXiv:hep-th/0509096].
  S.~Schafer-Nameki,
  ``Exact expressions for quantum corrections to spinning strings,''
  Phys.\ Lett.\  B {\bf 639}, 571 (2006)
  [arXiv:hep-th/0602214].
S.~Schafer-Nameki, M.~Zamaklar and K.~Zarembo,
  ``How accurate is the quantum string Bethe ansatz?,''
  JHEP {\bf 0612}, 020 (2006)
  [arXiv:hep-th/0610250].






 \bi{my}
  A.~A.~Tseytlin,
  ``Spinning strings and AdS/CFT duality,''
  hep-th/0311139.


 \bi{dol}
 F.~A.~Dolan and H.~Osborn,
  ``On short and semi-short representations for four dimensional superconformal
  symmetry,''
  Annals Phys.\  {\bf 307}, 41 (2003)
  [hep-th/0209056].
G.~Arutyunov and E.~Sokatchev,
  ``Conformal fields in the pp-wave limit,''
  JHEP {\bf 0208}, 014 (2002)
  [hep-th/0205270].


 \bi{krt}
 M.~Kruczenski and A.~A.~Tseytlin,
  ``Spiky strings, light-like Wilson loops and pp-wave anomaly,''
  Phys.\ Rev.\  D {\bf 77}, 126005 (2008)
  [0802.2039 [hep-th]].


 \bi{am2}
 L.~F.~Alday and J.~M.~Maldacena,
  ``Comments on operators with large spin,''
  JHEP {\bf 0711}, 019 (2007)
  [0708.0672 [hep-th]].

\bi{mtth}
R.~R.~Metsaev, C.~B.~Thorn and A.~A.~Tseytlin,
  ``Light-cone superstring in AdS space-time,''
  Nucl.\ Phys.\  B {\bf 596}, 151 (2001)
  [arXiv:hep-th/0009171].


\bi{afr}
G.~Arutyunov and S.~Frolov,
  ``Uniform light-cone gauge for strings in AdS(5) x S5: Solving $su(1|1)$
  sector,''
  JHEP {\bf 0601}, 055 (2006)
  [arXiv:hep-th/0510208].


\bibitem{Beccaria:2007cn}
  M.~Beccaria,
  ``Anomalous dimensions at twist-3 in the sl(2) sector of N = 4 SYM,''
  JHEP {\bf 0706}, 044 (2007)
  [arXiv:0704.3570 [hep-th]].



\bibitem{Beccaria:2007vh}
  M.~Beccaria,
  ``Universality of three gaugino anomalous dimensions in N = 4 SYM,''
  JHEP {\bf 0706}, 054 (2007)
  [arXiv:0705.0663 [hep-th]].

\bi{gvs}
N.~Gromov, S.~Schafer-Nameki and P.~Vieira,
``Efficient precision quantization in AdS/CFT,''
 arXiv:0807.4752 [hep-th]. \\
  N.~Gromov and P.~Vieira,
``The AdS(5) x S5 superstring quantum spectrum from the algebraic curve,''
  Nucl.\ Phys.\  B {\bf 789} (2008) 175
 [arXiv:hep-th/0703191].

\bibitem{fio}
  D.~Bombardelli, D.~Fioravanti and M.~Rossi,
  ``Large spin corrections in ${\cal N}=4$ SYM sl(2): still a linear integral
  equation,''
  arXiv:0802.0027 [hep-th].



\bi{iktt}
R. Ishizeki, M. Kruczenski, A. Tirziu and A.A. Tseytlin,
 Spiky strings in $AdS_3 \times S^1$ and their AdS-pp-wave limits,
 ArXiv:0812.2431.




\end{thebibliography}
\end{document}